\documentclass[aps,nofootinbib,preprint,showpacs,showkeys,tightenlines]{revtex4}
\usepackage{amsmath}
\usepackage{epsf}
\usepackage{epsfig}
\usepackage{axodraw}
\usepackage{amssymb}

\preprint{SNUTP 06-010} \preprint{KIAS-P06038}
\preprint{NSF-KITP-06-65}
\begin{document}
\title{\Large\bf Flipped $SU(5)$ from $Z_{12-I}$
orbifold  with Wilson line}
\author{Jihn E.
Kim$^{(a)}$\footnote{jekim@phyp.snu.ac.kr} and Bumseok
Kyae$^{(b)}$\footnote{bkyae@kias.re.kr} }
\address{
$^{(a)}$Department of Physics and Astronomy, and Center for
Theoretical Physics,\\
Seoul National University,
 Seoul 151-747, Korea,\ and\\
$^{(b)}$School of Physics, Korea Institute for Advanced Study,\\
207-43 Cheongryangri-dong, Dongdaemun-gu, Seoul 130-722, Korea}

\def\Qem{{$Q_{\rm em}$}}
 \def\pmhalf{{$\pm\frac12$}}
 \def\half{{$\frac12$}}
 \def\SMSSM{${\cal S}$MSSM\ }
\def\CPT{${\cal CPT}$\ }

 \def\N{{$\cal N$}}
 \def\Z{{\bf Z}}
 \def\MG{{$M_{\rm GUT}$}}

\begin{abstract}
 {We construct a three family flipped SU(5) model
from the heterotic string theory compactified on the $\Z_{12-I}$
orbifold with one Wilson line. The gauge group is $\rm SU(5)\times
U(1)_X\times U(1)^2\times[SU(2)\times SO(10)\times U(1)^2]^\prime$.
This model does not derive any nonabelian group except SU(5) from
$E_8$, which is possible only for two cases in case of one shift
$V$, one in ${\bf Z}_{12-I}$ and the other in ${\bf Z}_{12-II}$. We
present all possible Yukawa couplings. We place the third quark
family in the twisted sectors and two light quark families in the
untwisted sector. From the Yukawa couplings, the model provides the
$R$-parity, the doublet-triplet splitting, and one pair of Higgs
doublets. It is also shown that quark and lepton mixings are
possible.  So far we have not encountered a serious phenomenological
problem. There exist vectorlike flavor SU(5) exotics (including
\Qem=$\pm\frac16$ color exotics and \Qem=$\pm\frac12$
electromagnetic exotics) and SU(5) vectorlike singlet exotics with
\Qem=$\pm\frac12$ which can be removed near the GUT scale. In this
model, ${\rm sin}^2\theta_W^0=\frac{3}{8}$ at the full unification
scale.}
\end{abstract}

 \maketitle

\section{Introduction}

At present, it is of utmost importance to connect the high energy
string theory with the low energy standard model, in particular with
the minimal supersymmetric standard model (MSSM). The initial
attempt of the Calabi-Yau space compactification, which is
geometrical, has been very attractive \cite{candelas}. But the
orbifold compactification, also being a geometrical device, got more
interest due to its simplicity in model buildings
\cite{DHVW,inqsix}. Initially, the standard-like models were looked
for \cite{iknq}, in an attempt to obtain minimal supersymmetric
standard models (MSSMs), but it became clear that the standard-like
models have a serious problem on $\sin^2\theta_W$ to arrive at MSSMs
\cite{kim03}. All $\Z_{\rm N}$ models without Wilson lines were
tabulated a long time ago \cite{KobTab} and recently all $\Z_3$
models with Wilson lines are tabulated in a book \cite{ChoiKim05}.

The $\sin^2\theta_W$ problem is that it is better for the bare
value of $\sin^2\theta_W^0$ at the unification or string scale to
be close to $\frac38$ \cite{kim03} so that it reproduces the fact
of the convergence of three gauge couplings at one point near the
unification scale \cite{Langacker}. The so-called flipped SU(5)
does not fulfill this requirement automatically due to the leakage
of $\rm U(1)_Y$ beyond SU(5). Thus, the $\sin^2\theta_W$ problem
directs toward grand unified theories(GUTs) from superstring
without the electroweak hypercharge $Y$ leaking outside the GUT
group. In this regard, one may consider simple GUT groups  SU(5)
\cite{GG}, SO(10) \cite{SO10}, $\rm E_6$ \cite{E6} and
trinification  $\rm SU(3)^3$ \cite{tri}. The simplest orbifold
without matter representations beyond the fundamentals require the
Kac-Mooday level $K=1$. With $K=1$, one cannot obtain adjoint
representations \cite{ChoiKim05}; thus among the above GUT groups
the trinification group is the allowed one. Also, the Pati-Salam
$\rm SU(4)\times SU(2)\times SU(2)$ \cite{PatiSalam} can be broken
to the standard model without an adjoint matter representation;
above the GUT scale however three gauge couplings of the
Pati-Salam model diverge rather than evolving in unison. Thus,
trinification GUT seems to be the most attractive solution
regarding the $\sin^2\theta_W$ problem. The trinification is
possible only in ${\bf Z}_3$ orbifolds \cite{tristring}.

Another interesting GUT group, though not giving
$\sin^2\theta_W=\frac38$ necessarily, is the flipped SU(5)
\cite{antisu5,Deren} where the exchanges $d^c\leftrightarrow u^c$
and $e^c\leftrightarrow ({\rm neutral\ singlet}\ \nu^c)$ in the
representations of SU(5) are adopted. The matter representation of
the flipped-SU(5) is, under $\rm SU(5)\times
U(1)_X$,\footnote{Flavor SU(5) quantum number $X$ is  highlighted.}
\begin{equation}
{\bf 16}_{\rm flip}\equiv {\bf 10}_{\bf 1}+\overline{\bf 5}_{\bf
-3}+{\bf 1}_{\bf 5}
\end{equation}
The electroweak hypercharge is given by
\begin{equation}
Y=\textstyle\frac15 (X+Y_5)\label{Yflipped}
\end{equation}
where $Y_5={\rm diag.}(\frac13~\frac13~\frac13~-\frac12~-\frac12)$
and $X=\textstyle  {\rm diag.}(x~x~x~x~x)$. Then, the electroweak
hypercharges  of ${\bf 1}_{\bf 5}$ and $\overline{\bf 5}_{\bf -3}$
are +1, $-\frac23$, and $-\frac12$, which are $e^c, u^c,$ and
electron doublet. There are some nice features of flipped SU(5)
\cite{reffsu5}.

From the string context, flipped SU(5) was considered before in the
fermionic construction scheme \cite{Antoniadis:1989zy} and recently
in orbifold construction also \cite{SMSSM},  Calabi-Yau
compactification \cite{Blumenhagen:2006ux}, and intersecting D brane
models \cite{interD}. Let us call flipped SU(5) from string
construction `string flipped' SU(5). In string flipped SU(5), it
does not necessarily predict $\sin^2\theta_W=\frac38$ at the
unification scale \cite{kim03}. However, if we introduce more
parameters intrinsic in flipped SU(5), we may fit parameters so that
the gauge couplings meet at one point at the string scale $M_s$, the
unification scale of SU(5) and $\rm U(1)_X$ couplings. These
parameters include the symmetry breaking scale \MG\ for $\rm
SU(5)\times U(1)_X\to $ SM breaking and {\it intermediate scales of
vectorlike representations}. Above \MG\ the RG evolutions of SU(5)
and $\rm U(1)_X$ couplings are different, and we do not expect a
string scale around $0.7\times 10^{18}$ GeV \cite{kaplunovsky} but
can be determined by the unification ansatz at $M_s$ and mass scales
of vectorlike representations \cite{iwkim01}. In our string orbifold
model, ${\rm sin}^2\theta_W$ turns out to be $\frac{3}{8}$ at the
full unification scale due to the possibility of the electroweak
hypercharge embedding in SO(10). Thus such vectorlike fields should
be removed near the GUT scale.

String models in general include exotics. Electromagnetic exotics
(E-exotics) are fractionally charged particles which are nonabelian
gauge group singlets. Color exotics (C-exotics) are quarks with
non-standard charges, i.e. color triplet quarks not having
\Qem=$\frac23$ or $-\frac13$ and color anti-triplet quarks not
having \Qem=$-\frac23$ or $\frac13$. Flipped SU(5) GUT exotics
(G-exotics) are SU(5) representations, not having the $X$ charges of
${\bf 10}_{\bf 1}, \overline{\bf 10}_{\bf -1}, {\bf 5}_{\bf 3},
\overline{\bf 5}_{\bf -3}, {\bf 5}_{\bf -2}, \overline{\bf 5}_{\bf
2}, {\bf 1}_{\bf \pm 5}, {\bf 1}_{\bf 0}$. G-exotics contain
C-exotics and fractionally charged leptons.

The fermionic construction of flipped SU(5) has shown the existence
of E-exotics with \Qem=$\pm\frac12$ and integer charged `cryptons'
where cryptons are defined to be the composites of the hidden sector
confining group $\rm SU(4)^\prime$ \cite{crypton}. Cosmological
effect of cryptons was given in \cite{Crypcosmo}. Because of the
possibility that fractionally charged particles  exist in most
string vacua, discovery of fractionally charged particles may
strongly hint the correctness of the idea of string
compactification.

In this paper, we present the orbifold compactification with
$\Z_{12-I}$ twist. This contains a detailed account of
Ref.~\cite{SMSSM}. In addition, we present another orbifold model
having a hidden sector $\rm SU(4)^\prime$. We succeeded in
constructing a phenomenologically desirable flipped SU(5) model
from ${\bf Z}_{12-I}$.

We need three families of ${\bf 16}_{\rm flip}$, where
\begin{equation}
{\bf 16}_{\rm flip}\equiv {\bf 10}_{\bf 1}+\overline{\bf 5}_{\bf
-3}+{\bf 1}_{\bf 5}=(d^c,q,\nu^c)+(u^c,l)+e^c
\end{equation}
For spontaneous symmetry breaking, we need also the Higgs fields,
\begin{align}
({\bf 10}_{\bf 1}+\overline{\bf 10}_{\bf -1}) +({\bf 5}_{\bf
-2}+\overline{\bf 5}_2) ~.
\end{align}
%
%
%
Sometimes, it is useful to represent the components in terms of
\begin{equation}
{\bf 10}_{1}:\left(\begin{array}{cc}
 d^c & q \\ q & \nu^c
 \end{array}\right),\quad
\overline{\bf 5}_{\bf -3}:\left(\begin{array}{c}
 u^c \\ l
 \end{array}\right),\quad {\bf 1}_{5}:e^c,\quad
 {\bf 5}_{\bf -2}:\left(\begin{array}{c}
 D \\ h_d
 \end{array}\right),\quad
 \overline{\bf 5}_{2}:\left(\begin{array}{c}
 \overline{D} \\ h_u
 \end{array}\right),\label{flsu5rep}
\end{equation}
where $q$ and $l$ are lepton and quark doublets, $ D$ is
\Qem=$-\frac13$ quark, and $h_{d,u}$ are Higgs doublets giving mass
to $d,u$ quarks.
Spontaneous symmetry breaking of flipped SU(5) proceeds via VEVs of
${\bf 10}_{\bf 1}$ and $\overline{\bf 10}_{\bf -1}$ (components $
\langle\nu^c\rangle, {\langle\overline{\nu^c}\rangle}$) and ${\bf
5}_{\bf -2}$ and $\overline{\bf 5}_{\bf 2}$ (components in $\langle
h_d\rangle, \langle h_u\rangle$).

In this model, there appear two light families from the untwisted
sector and the third heavy family from twisted sectors. This is
dictated from the Yukawa coupling stucture. It also leads to
 (i) the doublet-triplet splitting, (ii) one pair of
Higgs doublets, and (iii) the existence of $R$-parity.

In Sec. \ref{sec:review}, we present a review on orbifold
construction with order $N=12$. Here we include formulae with Wilson
lines also.  In Sec. III and IV, the untwisted and twisted sector
spectra are calculated in detail. In Sec. V, we collect all
observable sector fields. In Sec. VI, we present the Yukawa coupling
structure and derive some phenomenological consequences. In Sec.
VII, we show ${\rm sin}^2\theta_W^0=\frac{3}{8}$. Sec. VIII is a
conclusion. In Appendix, we provide another model having $\rm
SU(4)'$.

\section{Orbifold method}\label{sec:review}

\begin{figure}[t]
\begin{center}
\begin{picture}(400,90)(0,-20)

\SetWidth{0.5} \Oval(74,24)(40,50)(0)\Curve{(45,35)(70,20)(95,35)}
\Curve{(50,30)(72,38)(90,30)}

\SetWidth{1.2} \Text(60,0)[c]{$\bullet$}

\SetWidth{0.5}
\Oval(189,24)(40,50)(0)\Curve{(160,35)(185,20)(210,35)}
\Curve{(165,30)(187,38)(205,30)}

\SetWidth{1.2} \Line(218.33,0)(214.34,-4) \Line(214.34,0)(218.34,-4)
\Text(174,0)[c]{$\bullet$} \GBox(202,18.4)(207,21.6){0}

\SetWidth{0.5}
\Oval(304,24)(40,50)(0)\Curve{(275,35)(300,20)(325,35)}
\Curve{(280,30)(302,38)(320,30)}

\SetWidth{1.2} \Text(288,0)[c]{$\bullet$}

\end{picture}
\caption{Simplified showing of fixed points of the ${\bf Z}_{12-I}$
orbifold.}\label{z12fixed}
\end{center}
\end{figure}
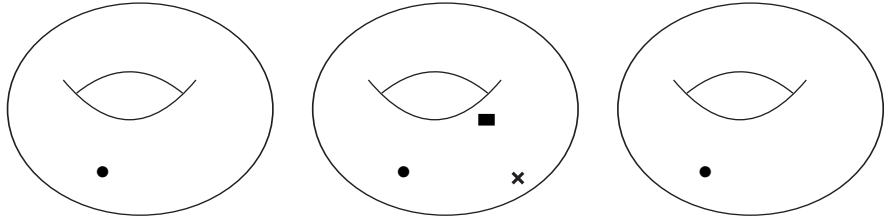

An orbifold is constructed from a manifold by identifying points
under a discrete symmetry group. The six internal space is
orbifolded by a twist vector $\phi_s$. With three complexified
components, $\phi_s$ has three components $\phi_{s1}, \phi_{s2}$ and
$\phi_{s3}$. The twist of $\Z_{12-I}$ orbifold is \cite{ChoiKim05},
\begin{eqnarray}
\textstyle \phi_s= (\frac{5}{12},~\frac{4}{12},~\frac{1}{12})\quad
{\rm with}\quad \phi_s^2=\frac{1}{12}\cdot \frac72.\label{twist}
\end{eqnarray}
Torus corresponding to $\phi_{si}$ is twisted by $\phi_{si}$. Hence
the first and third tori have one fixed point while the second torus
being modded by $\Z_3$ has three fixed points as schematically shown
in Fig. \ref{z12fixed}.
Multiplicity due to fixed points are three. These can be
distinguished by Wilson lines.

In ten dimensional (10D) heterotic string, left and right movers are
treated as gauge group degrees and \N=1 supersymmetry, respectively.
The embedding possibility is in the gauge group space of left
movers, NS sector of left and right movers, and R sector of right
movers. So we focus on the embedding in the group space for left
movers and in the R sector for right movers. In 10D, the R sector
embedding is given by $\phi_s$. The group space embedding is given
by sixteen numbers,  $V^I(I=1,2,\cdots,16) \equiv
\{v(I=1,\cdots,8),v^\prime(I=9,\cdots,16)\}$. Factoring out
$\frac1{12}$ by defining
$\phi_{sa}=\frac{1}{12}\phi_a,v_a=\frac{1}{12}w_a, v^\prime_a=
\frac{1}{12}w^\prime_a$, we must satisfy for a $\Z_{12-I}$ orbifold
\cite{AppD}
\begin{equation}\label{modular}
\begin{split}
&\sum \phi_a^2-\sum w^2_a -\sum w_a^{\prime 2}=0\ \ {\mbox{mod. 24}}\\
&3a_3=3a_4=0\ ,\quad a_1=a_2=a_5=a_6=0.\\
\end{split}
\end{equation}

\subsection{Dynkin diagram technique for finding gauge
group}\label{subsec:Dynkin}

\begin{figure}[t]
\begin{center}
\begin{picture}(400,80)(0,-20)

\SetWidth{0.5}
 \Oval(40,0)(7,7)(0) 
\Text(40,0)[c]{1} \Text(80,0)[c]{2}\Text(120,0)[c]{3}
\Text(160,0)[c]{4}\Text(200,0)[c]{5}
 {\SetWidth{1} \Line(90,10)(70,-10)\Line(90,-10)(70,10)}
 {\SetWidth{1} \Line(130,10)(110,-10)\Line(130,-10)(110,10)}
 \Text(240,0)[c]{6}
\Text(280,0)[c]{4}\Text(320,0)[c]{2}
 {\SetWidth{1}\Line(330,10)(310,-10)\Line(330,-10)(310,10)}
 {\SetWidth{1}\Line(290,10)(270,-10)\Line(290,-10)(270,10)}
 \CArc(80,0)(7,0,360)
\CArc(120,0)(7,0,360)
\CArc(160,0)(7,0,360)\CArc(200,0)(7,0,360)\CArc(240,0)(7,0,360)
\CArc(280,0)(7,0,360)\CArc(320,0)(7,0,360)
\Text(48,-13)[c]{$\alpha^0$}\Text(88,-13)[c]{$\alpha^1$}
\Text(128,-13)[c]{$\alpha^2$}
\Text(168,-13)[c]{$\alpha^3$}\Text(208,-13)[c]{$\alpha^4$}
\Text(248,-13)[c]{$\alpha^5$}
\Text(288,-13)[c]{$\alpha^6$}\Text(328,-13)[c]{$\alpha^7$}

\Line(47.5,0)(72.5,0) \Line(87.5,0)(112.5,0) \Line(127.5,0)(152.5,0)
\Line(167.5,0)(192.5,0)\Line(207.5,0)(232.5,0)
\Line(247.5,0)(272.5,0) \Line(287.5,0)(312.5,0)

\Text(240,40)[c]{3}
 {\SetWidth{1} \Line(50,10)(30,-10)
 \Line(50,-10)(30,10)}
  \CArc(240,40)(7,0,360) \Text(255,50)[c]{$\alpha^8$}
\Line(240,7.5)(240,32.5)

\end{picture}
\caption{The SU(5) subgroup of E$_8$. The Coxeter labels are shown
inside circles.}\label{e8tofsu5}
\end{center}
\end{figure}
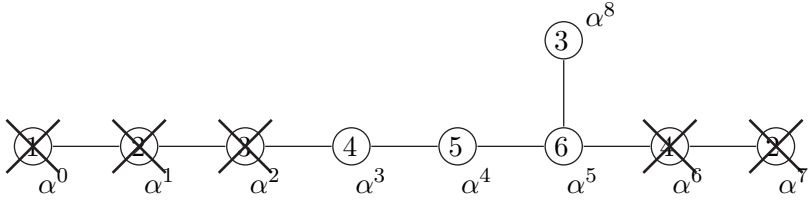

Just for finding out a gauge group structure, the Dynkin diagram
technique is extremely useful \cite{kacpet}. In the Dynkin diagram,
each simple root is endowed with a Coxeter label. A Dynkin diagram
technique of obtaining maximal subgroups is to strike out a simple
root from the extended Dynkin diagram. In orbifold, this is
generalized to strike out some roots where sum of the eliminated
Coxeter labels add up to order $N$ of $\Z_N$. To have SU(5) only
without a Wilson line, there must remain four linearly connected
simple roots. So, for $N\le 8$ orbifold, it is
impossible.\footnote{By a two step process using Wilson lines, it is
possible to obtain SU(5) in other $\Z_N$ orbifolds.} For $\Z_{12}$,
there is only possibility. Suppose $\sum_i c_i=N$ where $c_i$ is the
Coxeter label of simple root $\alpha_i$. There is only one
possibility which is $c_0+c_1+c_2+c_6+c_7=12$. See Fig.
\ref{e8tofsu5}. Thus, from orbifold compactification, there are only
two possibilities for constructing flipped SU(5) models, one in
$\Z_{12-I}$ and another in $\Z_{12-II}$. For $N=\sum_i c_i=N$, the
shift vector $V$ is given by  $V=\sum_i \Lambda_i$ where $\Lambda_i$
are fundamental weights \cite{AppD}. Thus, for $\Z_{12}$ orbifolds
$V$ is given by
$V_1=\Lambda_0+\Lambda_1+\Lambda_2+\Lambda_6+\Lambda_7$ with
$\Lambda_0=0$. Thus, the shift vector for flipped SU(5) is
$$
V_1=\textstyle\frac{1}{12}(\frac{17}{2}~\frac{5}{2}~\frac{3}{2}~
\frac{1}{2}~\frac{1}{2}~\frac{1}{2}~\frac{1}{2}~-\frac{1}{2})(\cdots).
$$
Now we shift the origin by $-\frac{5}{24}$, to obtain
\begin{equation}
V_2=\textstyle\frac{1}{12}(11~5~4~3~3~3~3~2 )(\cdots).\label{V2}
\end{equation}
Both $V_1$ and $V_2$ give an unbroken SU(5). But the entries of
$V_1$ and $V_2$ do not have five common entries. So, we try to add
an integer times $\Z_{12-I}$ shift $\phi_s$ so that the resulting
entries manifestly show five common entries. In this way, of course
the SU(5) nonabelian group is kept. Usually, if one adds $\phi_s$ to
three entries of $E_8$, some nonabelian groups are broken. So our
strategy is to add a multiple of $\phi_s$ such that an SU(5)
survives.  For this, we add
$(\frac{4}{12}~0~\frac{8}{12}~0~0~0~0~\frac{4}{12})$, to obtain
$\frac{1}{12}(15~5~12~3~3~3~3~6) $. Subtracting torus lattice and
rearranging entries, we obtain
\begin{equation}
V_3=\textstyle\frac{1}{12}(3~3~3~3~3~5~6~0)\label{V3}
\end{equation}
which has five common entries. This form is perfectly simple enough
in obtaining SU(5) weights since there are five common entries.
Otherwise, i.e. with $V_1$ or $V_2$, it is cumbersome to work out
all the SU(5) weights as tried out in \cite{ChoiKim05}.

Since the five entries are common, the simple roots for SU(5) take
the following standard form,
\begin{align}
\begin{array}{ccccccccccccc}
\alpha_{1} &= &( &1  &-1 &0 &0 &0 &; &0 &0 &0 &)\\
\alpha_{2} &= &( &0  &1 &-1 &0 &0&; &0 &0 &0 &)\\
\alpha_{3} &= &( &0  &0 &1 &-1 &0&; &0 &0 &0 &)\\
\alpha_{4} &= &( &0  &0 &0 &1 &-1&; &0 &0 &0 &)
\end{array}\label{simprootsu5}
\end{align}
Then, the highest weights of some representations we use are
\begin{eqnarray}
{\bf 5}: \left\{\begin{array}{cccccccccc}
(1&0&0&0&0&;0&0&0)\\
 (+\frac12&-\frac12&-\frac12&-\frac12&-\frac12&;0&0&0)
 \end{array}\right.\\
\overline{\bf 5}:\left\{ \begin{array}{cccccccccc}
(-1&0&0&0&0&;0&0&0)\\
(-\frac12&+\frac12&+\frac12&+\frac12&+\frac12&;0&0&0)
\end{array}\right.\\
{\bf 10}: \left\{\begin{array}{cccccccccc}
(1&1&0&0&0&;0&0&0)\\
(+\frac12&+\frac12&-\frac12&-\frac12&-\frac12&;0&0&0)
\end{array}\right.\\
\overline{\bf 10}: \left\{\begin{array}{cccccccccc}
(-1&-1&0&0&0&;0&0&0)\\
(-\frac12&-\frac12&+\frac12&+\frac12&+\frac12&;0&0&0)
\end{array}\right.
\end{eqnarray}

For the $\rm E_8^\prime$, we find roots and weights in a similar
way.

\subsection{Choosing shift vector and Wilson lines}

Embedding of the orbifold action can be found by satisfying the
modular invariance conditions (\ref{modular}). If one tries to have
a specific gauge group, the Dynkin diagram technique is very helpful
as we have discussed in Subsec. \ref{subsec:Dynkin}. Or one can
study the components of $(\cdots)(\cdots)^\prime$ to guess the gauge
group, but this method of finding gauge group is completed only
after one obtains all nonzero roots of the gauge multiplet. For
example, if one tries $V_2$ of (\ref{V2}) and $V_3$ of (\ref{V3})
then in both cases he will obtain SU(5). But guessing SU(5) from
(\ref{V2})is not straightforward. In this sense, the Dynkin diagram
technique is superior. On the other hand, a computer search of gauge
groups will cover all these cases. In the computer search, the
identical shift vector is given by several different forms as done
in $V_2$ and $V_3$. Usually, it is very difficult to identify all
the same shift vectors \cite{ChoiKim05}.

Choosing the shift vector ($V$) and Wilson lines ($a_1,\cdots,a_6$)
fixes the embedding of the orbifold action in the group space. So
gauge groups and representations are fixed by the shift vector and
Wilson lines, consistently with the modular invariance condition
(\ref{modular}).

In summary, the compactification is specified by twisting
represented by a shift form in the six internal space $\phi_s$,
shift $V$, and Wilson lines $a_1,\cdots,a_6$,
\begin{equation}
\begin{split}
&{\rm internal\ space}:\ \phi_s=(\phi_{s1},\phi_{s2},\phi_{s3})\\
&{\rm group\ space}:\ V,a_i(i=1,2,\cdots,6)
\end{split}
\end{equation}
Here, $s_0$ determines the chirality and $\tilde s$ encodes the
orbifolding information of the R sector.

\subsection{Massless modes}

Finding out all the massless modes below the compactification scale
is the key problem in the compactification process. The left movers
and right movers have different relations for the Einstein
mass-shell condition, even though the form has a similarity. In the
symmetric orbifold \cite{DHVW}, let us bosonize the Ramond sector of
right movers, which is represented by four half integers in $s$. The
orbifold action is modding out the 6D torus, and $s$ contains the
orbifold information of right movers under translation in the
internal space. For left movers, momenta $P$ corresponding to
translation in the group space have the orbifold information. All
these satisfy the level matching condition, $M_L^2=M_R^2$. Thus,
left moving and right moving states on torus have the following
vanishing vacuum energy for massless states,
\begin{equation}
\begin{split}
&{\rm left\ movers}:\ \frac{(P+kV)^2}{2}+\sum_jN^L_j\tilde{\phi}_j-\tilde c=0\\
&{\rm right\ movers}:\
\frac{(s+k\phi_s)^2}{2}+\sum_jN^R_j\tilde{\phi}_j-c=0,
\end{split}
\end{equation}
where $j$ runs over $\{1,2,3,\bar{1},\bar{2},\bar{3}\}$.  Here
$\tilde{\phi}_i\equiv k\phi_i$ mod. Z such that
$0<\tilde{\phi}_i\leq 0$, $\tilde{\phi}_{\bar{i}}\equiv -k\phi_i$
mod. Z such that $0<\tilde{\phi}_{\bar{i}}\leq 0$, and
$\phi_{s\bar{j}}\equiv\phi_{sj}$. (If $k\phi_i$ is an integer,
$\tilde{\phi}_j=1$~\cite{Kobayashi:2004ya,Buchmuller:2004hv}.)  For
$k=0$, conditions for massless left and right movers are given by
\begin{equation}
\begin{split}
& P^2=2 -2 \sum_jN^L_j\tilde{\phi}_j\\
& s^2=1-2\sum_jN^R_j\tilde{\phi}_j
\end{split}
\end{equation}
where $\tilde c=c+\frac12$.

The massless modes  include graviton $g_{\mu\nu}$, antisymmetric
tensor field $B_{\mu\nu}$, dilaton, gravitino, gauge bosons,
gauginos, and chiral matter. So the matter states $(P,s)$ must
satisfy
\begin{equation}
\begin{split}
& (P+kV)^2=2\tilde c-2 \sum_jN^L_j\tilde{\phi}_j \\
& (s+k\phi_s)^2=2c-2\sum_jN^R_j\tilde{\phi}_j\ .
\end{split}
\end{equation}
Among these massless modes, we are interested in a resulting SUSY
gauge theory. The SUSY condition for orbifold compactification is
given by right movers, which are given by $s$, i.e. $\tilde s$. We
introduce another set, $\tilde r$, of half integers so that $\tilde
s+\tilde r$ becomes integers,
\begin{equation}
\tilde r=(r_1,r_2,r_3).
\end{equation}
The SUSY condition for orbifolded spectrum is $\phi_s\cdot\tilde
r=0$, viz. Eq. (3.58) of \cite{ChoiKim05}.

\subsubsection{Gauge multiplet}

Gauge boson multiplets appear in the untwisted sector $U$,
satisfying
\begin{equation}
{\rm gauge\ group}:\ P^2=2,\ P\cdot V=0,\ P\cdot a_i=0\ {\rm for\
all\ }i\label{gaugeM}
\end{equation}
where the first one is the masslessness condition and the second
and the third ones are the orbifold conditions. The corresponding
right movers, satisfying the mass-shell and orbifold conditions
chooses two $s$'s with $s^2=1$, which are always \CPT\ conjugates
of each other. In this way, we obtain the gauge multiplet. As
expected, the multiplicity (${\cal P}$) of gauge bosons is 1.

\subsubsection{Matter multiplets}

Other massless states can appear in $U$ also, for
\begin{equation}
{\rm untwisted\ matter}:\ P^2=2,\ P\cdot V=\frac{k}{N}\ {\rm for\
}k\ne 0\ {\rm mod.\ }N,\ P\cdot a_i=0\ {\rm for\ all\
}i.\label{gaugeM}
\end{equation}
 We combine $\tilde s$ and $\tilde r$
to distinguish untwisted matter  so that they appear in three
categories under orbifolding, i.e. differing in sub-lattice
shifts,\footnote{For the same chirality, i.e. L, we use $(---)$
instead of $(+++)$. So, we take $U_1=(-1,0,0)$.}
\begin{equation}
U_1:\tilde s+\tilde r=(-1,0,0),\ U_2:\tilde s+\tilde r=(0,1,0),\
U_3:\tilde s+\tilde r=(0,0,1).\label{Ui}
\end{equation}

There are fixed points in field theory orbifolds. In string
orbifolds also, we must consider physics related to fixed points.
Massless strings can sit at fixed points, which is found by the
mass-shell condition at fixed point. But noting that some linear
combinations of strings sitting at several fixed points may be taken
to satisfy the orbifold condition, we consider twisted sectors. For
${\bf Z}_N$ orbifold, we consider $k=1,\cdots,N-1$ twisted sectors,
$T_k$. The \CPT conjugates of $T_k$ appear in $T_{N-k}$. Thus, in
non-prime orbifolds ${\bf Z}_4,{\bf Z}_6,{\bf Z}_8,{\bf Z}_{12}$,
the sector $T_{N/2}$ contains \CPT conjugates also.

For untwisted matter, multiplicity ($\cal P$) is given just by
counting all possible states. Multiplicity in the twisted sector
is more involved. The method of linear combination can come with
complex numbers. This is taken into account in the (generalized)
GSO projector, which projects out non-physical states. It can be
read off from the one loop partition function of
string~\cite{Ibanez:1987pj,ChoiKim05};
\begin{eqnarray} \label{projP}
{\cal
P}_k=\frac{1}{N}\sum_{l=0}^{N-1}\tilde{\chi}(\theta^k,\theta^l)~e^{2\pi
il\Theta_0} \equiv \frac{1}{N}\sum_{l=0}^{N-1}\tilde
\chi(\theta^k,\theta^l)\Delta^l~,
\end{eqnarray}
where $N$ is the order of ${\bf Z}_{N}$ orbifolds, and
\begin{eqnarray} \label{phase0}
\Theta_0 =
\sum_{j}\left(N^L_j-N^R_j\right)\hat{\phi}_j-\frac{k}{2}(V^2-\phi_s^2)
+ (P+kV)\cdot V-(\tilde s+k\phi_s)\cdot{\phi_s} +~{\rm integer}
\end{eqnarray}
where $j$ denotes the coordinates of the 6 dimensional compactified
space running over $\{1,2,3,\bar{1},\bar{2},\bar{3}\}$ in
complexified coordinates, and $\hat{\phi}_i=\phi_{si}{\rm
sgn}(\tilde\phi_i)$ where ${\rm
sgn}(\tilde\phi_{\overline{i}})=-{\rm sgn}(\tilde\phi_i)$
\cite{{Kobayashi:2004ya}}. The $\tilde{\chi}(\theta^m,\theta^k)$ in
Eq.~(\ref{projP}) denotes the degenerate factor tabulated in
Table~\ref{tb:degn}~\cite{ChoiKim05,chi2}. For the sectors wound by
Wilson lines, the Wilson line modified shift vector $V_f$ is used
instead of $V$.


In the presence of a Wilson lines ($\equiv a^I$), the GSO
projector Eq.~(\ref{projP}) needs to be modified
as~\cite{Kobayashi:2004ya}
\begin{eqnarray}\label{newprojP}
e^{2\pi il\Theta_0}\longrightarrow
\frac{1}{N_W}\sum_{f=0}^{N_W-1}e^{2\pi il\Theta_f} ~,
\end{eqnarray}
where $N_W=3$ for Wilson line of order three in case of ${\bf
Z}_{12-I}$, and $\Theta$ in Eq.~(\ref{phase0}) should also be
modified as
\begin{eqnarray}\label{phasef}
\Theta_f =
\sum_{j}\left(N^L_j-N^R_j\right)\hat{\phi}_j-\frac{k}{2}\left(V_f^2-\phi_s^2\right)+
(P+kV_f)\cdot V_f -(\tilde s+k\phi_s)\cdot{\phi_s}
 + ~{\rm integer} ~,
\end{eqnarray}
where $V_f\equiv (V+m_fa)$.

\begin{table}
\begin{center}
\begin{tabular}{c|cccccccccccc}
\hline 
 $k\diagdown l$ & $0$ & $1$ & $2$ & $3$ & $4$ & $5$ &
$6$ & $7$ & $8$ & $9$ & $10$ & $11$
\\
\hline $1$ & $3$ & $3$ & $3$ & $3$ & $3$ & $3$ & $3$ & $3$ & $3$ &
$3$ & $3$ & $3$
\\
$2$ & $3$ & $3$ & $3$ & $3$ & $3$ & $3$ & $3$ & $3$ & $3$ & $3$ &
$3$ & $3$
\\
$3$ & $4$ & $1$ & $1$ & $4$ & $1$ & $1$ & $4$ & $1$ & $1$ & $4$ &
$1$ & $1$
\\
$4$ & $27$ & $3$ & $3$ & $3$ & $27$ & $3$ & $3$ & $3$ & $27$ & $3$
& $3$ & $3$
\\
$5$ & $3$ & $3$ & $3$ & $3$ & $3$ & $3$ & $3$ & $3$ & $3$ & $3$ &
$3$ & $3$
\\
$6$ & $16$ & $1$ & $1$ & $4$ & $1$ & $1$ & $16$ & $1$ & $1$ & $4$
& $1$ & $1$
\\
\hline
\end{tabular}
\end{center}
\caption{Degeneracy factor $\tilde\chi(\theta^k,\theta^l)$ in the
${\bf Z}_{12-I}$ orbifold.}\label{tb:degn}
\end{table}

\section{$\Z_{12-I}$  model}

We choose the following shift vector and Wilson lines
\begin{align}
&V=\textstyle\left( \frac14~ \frac14~ \frac14~ \frac{1}{4}~ \frac14~
\frac{5}{12}~\frac{6}{12}~  0~ \right)\left( \frac{2}{12}~
\frac{2}{12}~ 0~ 0~0~0~0~0 \right)\label{shiftV}\\
&a_3=a_4=\textstyle\left( 0^5~0~ \frac{-1}{3}~ \frac{1}{3}~
\right)\left( 0~ 0~\frac{2}{3}~0^5 \right)\label{Wilson3}\\
&a_1=a_2=a_5=a_6=0\nonumber
\end{align}
which  satisfies Eq.~(\ref{modular}):  it gives
$V^2-\phi_s^2=\frac12$ modular $2\cdot(\rm integer)/12$. Since the
Wilson line is a ${\bf Z}_3$ shift, it distinguishes three cases.
These satisfy the following conditions
~\cite{Ibanez:1987pj,Forste:2004ie}:
\begin{equation}
\begin{split}
& 12(|V|^2-|{\phi}_s|^2) =0 ~~{\rm mod.~even~integer},
\\
& 12(V\cdot a_3)=0 ~~{\rm mod.~integer},
\\
& 12|a_3|^2 =0 ~~{\rm mod.~even~integer}.
\end{split}\label{WilCond}
\end{equation}
Thus, we consider  the following effective shifts distinguishing
twisted sector $T_k$ to $T_k^0,T_k^+$, and $T_k^-\ (k\ne 0,3,6,9)$
by
\begin{align}
&V_0\equiv V\\
 &V_+\equiv V+a_3=\textstyle\left( \frac14~ \frac14~
\frac14~ \frac{1}{4}~ \frac14~ \frac{5}{12}~ \frac{2}{12}~
\frac{4}{12}\right)\left( \frac{2}{12}~
\frac{2}{12}~\frac{8}{12}~0^5 \right)\\
&V_-\equiv V-a_3=\textstyle\left( \frac14~ \frac14~ \frac14~
\frac{1}{4}~ \frac14~ \frac{5}{12}~ \frac{10}{12}~
\frac{-4}{12}\right)\left( \frac{2}{12}~
\frac{2}{12}~\frac{-8}{12}~0^5 \right).
\end{align}

For future convenience, we list
\begin{equation}
\begin{split}
&\textstyle V_0^2-\phi_s^2=\frac12 ~,
\\
&\textstyle V_+^2-\phi_s^2=\frac56 ~,
\\
&\textstyle  V_-^2-\phi_s^2=\frac{3}{2}.
\end{split}
\end{equation}

\subsection{Gauge group from untwisted sector}

For the gauge multiplet, we search for roots satisfying $P^2=2,
P\cdot V=0$ and $P\cdot a_3=0$, and obtain the following unbroken
gauge group
\begin{equation}
[SU(5)\times U(1)_X\times U(1)^3]\times [SU(2)\times SO(10)\times
 U(1)^2]^\prime
\end{equation}
where we choose the $U(1)_X$ of flipped-SU(5) as
\begin{eqnarray}
Q_X=(-2,-2,-2,-2,-2;0,0,0)(0^8)'.
\end{eqnarray}
For example, one can see that the following is the roots of SU(5)
\begin{eqnarray}
 (\underline{1~ -1~~0~~0~~0}~;0~~0~~0)\quad
 &\quad\text{nonzero roots among }
 {\bf 24}\text{ of $SU(5)$}\label{tb2:z12}
\end{eqnarray}
where the underlined entries allow permutations.

\subsection{Matter from untwisted sector}
\subsubsection{\bf Chirality}
\def\hf{\frac12}
The chirality is determined by the 8 component SO(8) spinor of the
Ramond sector of right movers. It is labeled by $s=(s_0,\tilde
s)=\{\pm\hf,\pm\hf, \pm\hf,\pm\hf\}$ with an even number of minus
signs. We define the 4D chirality $\chi$ as the one originating from
the first entry of $s$ denoted by $\oplus$ or $\ominus$, i.e.
\begin{align}
&s=\textstyle\{\oplus\ {\rm or}\ \ominus,\pm\hf,
\pm\hf,\pm\hf\}\equiv\{\hf\chi,\tilde s\}\\
 &\chi=2s_0={2\oplus\ {\rm or}\ 2\ominus} ~.
\end{align}
Let us call $\chi=1(-1)$ as `right-(left-)handed', and $\tilde s$
has three components in terms of $\pm\hf$. In the untwisted sector
$U$, we have $U_1, U_2$ and $U_3$ as defined in Eq. (\ref{Ui}).
These $U_i$ correspond to the untwisted $\tilde
s=(---),(++-),(+-+)$, respectively [or to $(+++),(--+),(-+-)$,
respectively, since antiparticles can be used also].

The chirality in the twisted sector can be similarly defined by
the 8 component SO(8) spinor of the Ramond sector of right movers.

\subsubsection{\bf Spectrum}
 The massless matter fields are those with $P\cdot V\ne 0$ and
satisfy the masslessness condition.  $P$ must satisfy  $P^2=2,\
P\cdot a_3={\rm integer}$ and $P\cdot V=\frac{k}{12}$ where
$k=1,2,\cdots,11$.
 We will consider only $k
=1,2,\cdots,6$, since the rest will provide their ${\cal CTP}$
conjugates. Here, it is sufficient to look at 3 cases only,
$k=1,4,5$, to keep the GSO allowed states.

We have the convention that the highest weight of the complex
conjugated representation is in fact the lowest one so that all the
weights of the complex conjugated representation are obtained by
adding simple roots.

Let $\alpha$ denote the phase $e^{2\pi i/12}$. For $k=1$ or $P\cdot
V=\frac{1}{12}$, the left-movers obtain a phase $\alpha$. We need an
extra phase $\alpha^{-1}$ from the right movers, which is
accomplished by $e^{-2\pi is\cdot\phi_s}$ where
$\phi_s=\frac{1}{12}(5~4~1)$ and $s=(\ominus +-+)$. It is
left-handed and allows $U_3$. For $k=4$,  $s=(\ominus ++-)$ provides
the needed $\alpha^{-4}$; thus it is left-handed and provides $U_2$.
For $k=5$, $s=(\oplus +++)$ provides $\alpha^{-5}$; thus it
right-handed and provides $U_1$. $\alpha$ from the right movers is
provided by $s=(\oplus -+-)$ which thus will couple to $k=11$. It is
right-handed. $\alpha^{4}$ from the right movers is provided by
$s=(\oplus --+)$ which will couple to $k=8$. It is right-handed.
$\alpha^{5}$ from the right movers is provided by $s=(\ominus +--)$
which will couple to $k=7$. It is left-handed. These particles for
$k>6$ give the antiparticle spectra. The chiralities and $U_i$s are
shown in Tables \ref{tb:untwistedvi} and \ref{tb:untwistedhd}.

\begin{table}
\begin{center}
\begin{tabular}{|c|c|c|c|}
\hline $P\cdot V$& $\tilde s,\ U_i$ & Visible states & $SU(5)\times
U(1)_X$
\\
\hline  & &$(\underline{+----};+++)$ & ${\bf {5}}_{\bf 3}^{L}$

\\
$\frac{1}{12}$ & $(+~-~+),\ U_3$ &$(\underline{+++--};+--)$ &
$\overline{\bf 10}_{\bf -1}^{L}$
\\
$$ &  &$(\underline{+++++};+++)$ & ${\bf 1}_{\bf -5}^{L}$
\\[0.2em]
\hline $\frac{4}{12}$ & $(+~+~-),\ U_2$
&$(\underline{-1,0,0,0,0};-1,0,0)$ & $\overline{\bf 5}_{\bf 2}^{L}$
\\[0.2em]
\hline &  & $(\underline{++++-};-++)$ & $\overline{\bf {5}}_{\bf
-3}^{R}$
\\
$\frac{5}{12}$&  $(+~+~+),\ U_1$ & $(\underline{++---};---)$ & ${\bf
10}_{\bf 1}^{R}$
\\
$$ & &$(-----;-++)$ & ${\bf 1}_{\bf 5}^{R}$
\\
\hline
\end{tabular}
\end{center}
\caption{Visible sector chiral fields from the
$U$-sector.}\label{tb:untwistedvi}
\end{table}

\begin{table}
\begin{center}
\begin{tabular}{|c|c|c|c|c|}
\hline $P\cdot V$ & $\tilde s,\ U_i$ & Hidden states &
$[SU(2)\times SO(10)]'$ & Label
\\
\hline $\frac{4}{12}$& $(+~+~-),\ U_2$ & $(1,1;0,0;0,0,0,0)'$ &
${\bf 1_0}$ & $s^u$
\\[0.2em]
\hline
\end{tabular}
\end{center}
\caption{Hidden sector chiral fields from the
$U$-sector.}\label{tb:untwistedhd}
\end{table}

Thus, the vectors for $p\cdot V=\frac{1}{12}, \frac{5}{12}$, and
$\frac{4}{12}$ give,
\begin{equation}
({\bf 1}_{\bf -5}+{\bf {5}}_{\bf 3}+\overline{\bf 10}_{\bf
-1})^L_{U_3},\quad (\overline{\bf 5}_{\bf 2})_{U_2}^{L},\quad({\bf
1}_{\bf -5}+{\bf {5}}_{\bf 3}+\overline{\bf 10}_{\bf
-1})^L_{U_1},\quad ({\bf 1_0})^L_{U_2}\label{specunvis}
\end{equation}
plus ${\cal CTP}$ conjugates. Here we listed only $U(1)_X$ charges
as boldfaced subscripts. Note that for the untwisted $k=6$ sector
there is no way to provide an additional $\alpha^6$ at the massless
level from the right movers.

\section{Twisted sectors with Wilson line}

The $Z_{12-I}$ with the twist vector $\phi=\frac{1}{12}(5,4,1)$ has
three fixed points in the second torus for  the prime order
$\theta^1$ and $\theta^5$ twists. This is because it is the same as
$\Z_3$. For the first and the third torus the origin is the only
fixed point, viz. Fig. \ref{z12fixed}. For the other twists such as
$k=4$ and $6$, counting the number of massless states involves a
more complicated nonvanishing projection operator $P_{\theta^k}$.
But addition of Wilson lines can distinguish some fixed points. In
fact, for $Z_{12-I}$ possible Wilson lines must satisfy $3a_3=0,
3a_4=0$ and $a_i=0\ (i\ne 3,4)$ so that any combination of
$k(V+m_fa_i)$ is another shift vector. Since $\Z_{12-I}$ allows
$3a_3=0$ modular integer, the Wilson line is a $\Z_3$ shift.

In the second torus, Wilson lines must be symmetric, $a_3=a_4$.
Then, the $k^{\rm th}$ twisted sector is distinguished by $kV,
k(V+a_3)$, and $k(V-a_3)$, which is denoted as $\tilde V$,
\begin{equation}
\tilde V=k(V+m_fa_3)\ ,\quad m_f=0,\pm 1
\end{equation}
or
\begin{equation}
\tilde V=k\{V_0,\ V_+,\ V_-\}.
\end{equation}
Since $3a_3=6a_3=0$, $2V_{\pm}$ and $5V_{\pm}$ in the $T_2$ and
$T_5$ sectors are equivalent to $2V\mp a_3$ and $5V\mp a_3$,
respectively. On the other hand, $4V_{\pm}$ in the $T_4$ sector is
equivalent to $4V\pm a_3$.

The masslessness condition is
\begin{equation}
(P+\tilde V)^2=2(1-c_k)-2\tilde N_L.\label{condit3}
\end{equation}
For the $\theta^k$ twist$(k=1,2,\cdots,6)$, we have
\begin{equation}
2(1-c_k)^L=\left\{\begin{array}{cccc} \frac{210}{144},\ & k=1;&
\quad\ \frac{192}{144},\ & k=4\\
\frac{216}{144},\ & k=2; &\quad\ \frac{210}{144},\ & k=5\\
\frac{234}{144},\ & k=3; &\quad\ \frac{216}{144},\ & k=6,
\end{array} \right.\label{vacener1}
\end{equation}
for the left movers, and
\begin{equation}
2(1-c_k)^R=\left\{\begin{array}{cccc} \frac{11}{24},\ & k=1;&
\quad\ \frac{1}{3},\ & k=4\\
\frac{1}{2},\ & k=2; &\quad\ \frac{11}{24},\ & k=5\\
\frac{5}{8},\ & k=3; &\quad\ \frac{1}{2},\ & k=6.
\end{array} \right. \label{vacenerR}
\end{equation}
for the right movers.

For the sectors wound by Wilson lines, $V_\pm$ are used instead of
$V_0$. The untwisted sector $k=0$ and twisted sectors for $k=3,6,9$
are not affected by Wilson lines since the Wilson line condition,
$3a_3=0$, makes it trivial. So, for $ k=3,6,9$, there is the
additional condition, $ (P+kV)\cdot a_i=0$, which is applicable to
$T_6$ only in our case. For $k\ne 3,6,9$, the multiplicity for each
twisted sector $k(V+m_f a_3)$ is ${\cal P} = \frac13 {\cal P}_k$.

The formula for multiplicity, Eq. (\ref{projP}), is the GSO
allowed number of states.\footnote{For the prime orbifold $\Z_3$,
the multiplicity is just $\frac13(1+\Delta+\Delta^2)$ which can be
either 1 for $\Delta=1$ or 0 for  $\Delta=e^{\pm 2\pi i/3}$. So in
$\Z_3$ it is sufficient to count those with the vanishing phase,
i.e. $(P+V)\cdot V-(s+\phi)\cdot\phi=0$. It is so also in the
$k$th twisted sector of ${\bf Z}_{12-I}$ if $\tilde{\chi}(k,l)$
are the same for all $l$.} For nonprime orbifolds such as
$\Z_{12}$, the multiplicity (\ref{projP}) is nonvanishing even if
$\Delta$ were not 1. Only for those with pure $\Z_{12}$ twists,
i.e. $k=1,~2$ and 5, the multiplicity is counted by those with the
vanishing phase.

The twisted sectors for $k=3,6,9$ are not affected by the
additional Wilson lines since $3a_3=0$. Note that the untwisted
sector also is not distinguished by Wilson lines, but  Wilson
lines give the modular invariance condition $P\cdot a_i=0$ in  the
untwisted sector. By the same token, in the sectors where $3a_3=0\
(k=0, 3,6,9,\ {\rm with\ 0\ corresponding\ to\ the\ untwisted\
sector})$, the modular invariance condition restricts Wilson
lines~\cite{Kobayashi:2004ud},
\begin{equation}
(P+kV)\cdot a_3=0,\quad k=0,3,6,9.\label{Wilsontw}
\end{equation}
But other twisted sectors are affected, in particular the
$k=1,2,4,5$ sectors. For these sectors, the multiplicity for each
of $k(V+m_f a_3)$ is $\frac13 P_{\theta^k}$ where $P_{\theta^k}$
is given in Eq. ({\ref{projP}}).

Let us present twisted sectors for $k=6,1,2,3,4,5$ in order. The
$k=6$ twisted sector $T_6$ contains the ${\cal CTP}$ conjugates in
$T_6$ again. The spectra in the $k=1,\cdots,5$ twisted sectors
accompany their ${\cal CTP}$ conjugates in $k=7,\cdots,11$.

\subsection{Twisted sector $T_6$}

The massless condition for the left mover is
$\frac{1}{2}|P+6V|^2+\sum_j(N^L)_j(\tilde{\phi})_j=\frac{3}{4}$,
where $P$ is the $E_8\times E_8'$ weight vectors.
The left mover states satisfying the massless condition always
appear vector-like in the $T_6$-sector. In general, however, they
carry different phases from those of the counterpart states with
opposite quantum numbers. Thus, chiral matter spectrum is possible
even in $T_6$ after imposing the GSO projection by
Eq.~(\ref{projP}).
In view of (\ref{Wilsontw}), we additionally require
\begin{eqnarray}
(P+6V)\cdot a_3=0~~{\rm mod.~Z} ~.
\end{eqnarray}
 The massless states with left handed chirality
satisfying this constraint, and their multiplicity numbers
determined by ${\cal P}_6$ are listed in Table \ref{T6states}.

\begin{table}
\begin{center}
\begin{tabular}{|c|c|c|c|c|c|}
\hline  $P+6V$ & $(N^L)_j$ & $\Theta_0$ & ${\cal P}_6$& $\chi$ &
Labels
\\
\hline
 $\left({\bf 5}_{\bf 3};\frac{1}{2},0,0\right)(0^8)'$ & $0$
 & $\frac{1}{2}$ & $2$
&$L$ &
\\
 $\left(\overline{{\bf 10}}_{\bf -1};\frac{1}{2},0,0\right)(0^8)'$
& $0$ & $0$ & $4$ &$L$ &
\\
$\left({\bf 1}_{\bf -5};\frac{1}{2},0,0\right)(0^8)'$ & $0$ &
$\frac{1}{2}$ & $2$ &$L$ &
\\[0.2em]
\hline
 $\left(\overline{{\bf 5}}_{\bf -3};\frac{-1}{2},0,0\right)(0^8)'$ & $0$ &
 $\frac{-1}{6}$ & $2$ &$L$ &
\\
$\left({\bf 10}_{\bf 1};\frac{-1}{2},0,0\right)(0^8)'$ & $0$ &
$\frac{1}{3}$ & $3$ &$L$ &
\\
 $\left({\bf{1}}_{\bf 5};\frac{-1}{2},0,0\right)(0^8)'$ & $0$ &
$\frac{-1}{6}$ & $2$ &$L$ &
\\[0.2em]
 \hline
$$ & $1_3$ & $0$ & $4$ &$L$ & $h_1$
\\
$\left({\bf 1}_{\bf 0};0,\frac{1}{2},\frac{1}{2}\right)(0^8)'$ &
$1_{\bar{3}}$ & $\frac{-1}{6}$ & $2$ &$L$ & $h_2$
\\
$$ & $1_{\bar{1}}$ & $\frac{1}{2}$ & $2$ &$L$ & $h_3$
\\
$$ & $1_{1}$ & $\frac{1}{3}$ & $3$ &$L$ & $h_4$
\\[0.1em]
\hline
$$ & $1_3$ & $\frac{1}{2}$ & $2$ &$L$ & $\bar{h}_1$
\\
$\left({\bf 1}_{\bf 0};0,\frac{-1}{2},\frac{-1}{2}\right)(0^8)'$ &
$1_{\bar{3}}$ & $\frac{1}{3}$ & $3$ &$L$ & $\bar{h}_2$
\\
$$ & $1_{\bar{1}}$ & $0$ & $4$ &$L$ & $\bar{h}_3$
\\
$$ & $1_{1}$ & $\frac{-1}{6}$ & $2$ &$L$ & $\bar{h}_4$
\\[0.2em]
\hline
\end{tabular}
\end{center}
\caption{Left-handed massless states satisfying $(P+6V)\cdot a_3=0 $
mod. Z in $T_6$.} \label{T6states}
\end{table}

For simplicity, here we employed the following abbreviations for
the SO(10) spinors and neutral singlets under $\rm SU(5)\times
U(1)_X$,
\begin{align}
&{\bf {5}}_{\bf 3}\equiv (\underline{+----})~, ~~\overline{\bf
10}_{\bf -1}\equiv (\underline{+++--})~,~~{\bf 1}_{\bf -5}\equiv
(+++++)
\\
&\overline{\bf 5}_{\bf -3}\equiv (\underline{-++++})~, ~~{\bf
{10}}_{\bf 1}\equiv (\underline{---++})~,~~{\bf {1}}_{\bf 5}\equiv
(-----) ~,
\\
&{\bf 1_0}\equiv (0,0,0,0,0) ~.
\end{align}

From the $T_6$-sector, thus we have the following massless states,
\begin{eqnarray}
\begin{split}
& {\bf 10}_{\bf 1}^R +3\{ {\bf 10}_{\bf
1}^R,~\overline{\bf{10}}_{\bf -1}^R\} +2 \{\overline{\bf {5}}_{\bf
-3}^L,~{\bf 5}_{\bf 3}^L\} + 2 \{{\bf 1}_{\bf
5}^L,~ {\bf {1}}_{\bf -5}^L\} +22\ {\rm neutral~singlets}\\
&+{\cal CTP}{\rm ~conjugates.}
\end{split}
\end{eqnarray}
While the multiplicity number for $\overline{\bf{10}}_{-1}$ is 3,
the multiplicity of ${\bf 10}_{\bf 1}$ is 4.  One of ${\bf 10}_{\bf
1}^R$s provides one generation of the MSSM matter, $\{Q$, $d^c$,
$\nu^c\}$. The remaining vector-like pairs of $\{{\bf 10}_{\bf 1},
~{\bf\overline{10}}_{-1}\}$ could be utilized to break $SU(5)\times
U(1)_X$ into the MSSM gauge group.

We present the calculation in detail for the first row. The
$\theta^6$ twist vectors are
\begin{equation}
\begin{split}
&\tilde \phi= 6 \phi_s=\textstyle(\frac 12,0,\frac12)\\
&\tilde V= 6V\equiv
\textstyle(\frac12~\frac12~\frac12~\frac12~\frac12~\frac12~0~0)(0~0~
0~0~0~0~0~0).
\end{split}\label{T6vec}
\end{equation}
In the $\theta^6$ twisted sector, consider $P=(0^8)(0^8)$. With some
shifts, $P$ can be $([-1]^5;-2,-3,0)(-1,-1;-6,0^5)^\prime.$ Then,
from (\ref{condit3}) and (\ref{vacener1}) we require
$(P+6V)^2=\frac32\to 2\tilde c =\frac{216}{144}=\frac32$ which is
certainly satisfied with $\tilde N_L=0$. Therefore, the state
$P=(0^8)(0^8)$ is massless.
 The GSO projection is given when combined with
the right movers. The masslessness condition for right movers is
$(s+6\phi)^2=\frac12$,\footnote{See Appendix D of \cite{ChoiKim05}.}
from which we will determine the chirality.

Let us calculate the masslessness condition of the left movers
first, the multiplicity and the chirality.

The masslessness condition for left movers becomes
\begin{equation}\label{theta6}
(P+\tilde V)^2+2\sum_jN^L_j\tilde{\phi}_j=\frac32.
\end{equation}
Thus, the vectors satisfying Eq. (\ref{theta6}) with
Eq.~(\ref{T6vec}) constitute the representation ${\bf \bar 5}_L$.
These weights are
\begin{equation}
\begin{split}
{\bf 1}_{\bf 5}:\quad &({0~0~0~0~0}~0~0~0)\\
{\overline{\bf 5}}_{\bf -3}:\quad&(\underline{-1~0~0~0~0}~-1~0~0)\quad\\
{\bf 10}_{\bf 1}:\quad&(\underline{-1~-1~-1~0~0}~-1~0~0)
\end{split}
\end{equation}
where we calculated $U(1)_X$ charges as $(P+\tilde V)\cdot Q_X$.
Their ${\cal CTP}$ conjugates are
\begin{equation}
\begin{split}
 &({-1~-1~-1~-1~-1}~-1~0~0)\\
&(\underline{-1~-1~-1~-1~0}~0~0~0)\quad\\
&(\underline{-1~-1~0~0~0}~0~0~0)
\end{split}
\end{equation}

Note that for $P=(0^8)(0^8)$,
\begin{equation}
(P+6V)\cdot V=\textstyle \frac{5}{6},\ \ \tilde \phi\cdot
\phi_s=\textstyle \frac14.
\end{equation}
To obtain the chirality $\chi$, we look at $s=(\frac12\chi,\tilde
s)$ allowing nonvanishing multiplicities.  Then the phase of
$\Delta_{\theta^6}$ is found as $(P+\tilde V)\cdot V-(\tilde s+
\tilde \phi)\cdot \phi_s-\frac12 k(V^2-\phi_s^2)$. With $k=6$, we
have $\frac12 k(V^2-\phi_s^2)=\frac32=\frac12$, and the phase of
$\Delta$ is $2\pi$ times $(\frac{1}{12}-\tilde s\cdot\phi_s)$, and
hence we obtain the multiplicity listed in Table \ref{tbmult}. Then,
we read the chirality $\chi$, or $\oplus$ or $\ominus$, from the
first entry of $s=(\frac12\chi|s_1~s_2~s_3)$ if that chirality is
allowed by the right mover condition, which is shown in the first
column of Table \ref{tbmult}. If the first column does not satisfy
the right mover condition, we should search for higher $s_0$, which
will be discussed shortly.
\begin{table}
\begin{center}
\begin{tabular}{|c|c|c|c|c|}
\hline &&&&\\
[-1em] $\frac12\chi$& $\tilde s=(r+\tilde\omega)$ & $\tilde
s\cdot\phi_s$ &$\Delta$ phase& Multiplicity\\
[0.2em] \hline
&&&&\\[-1em]
$\oplus$ & $~+~+~+~$ & $\frac{5}{12}$
&$-\frac{4}{12}\cdot 2\pi $ &3\\[0.3em]
$\oplus$ & $~+~-~-~$ & $0$
&$\frac{1}{12}\cdot 2\pi $ &0\\[0.3em]
$\oplus$ & $~-~+~-~$& $-\frac{1}{12}$
&$\frac{2}{12}\cdot 2\pi  $ &2\\[0.3em]
$\oplus$ & $~-~-~+~$& $-\frac{4}{12}$
&$\frac{5}{12}\cdot 2\pi  $ &0\\[0.3em]
$\ominus$ & $~-~-~-~$ & $-\frac{5}{12}$
&$\frac{6}{12}\cdot 2\pi $ &2\\[0.3em]
$\ominus$ & $~-~+~+~$ & $0$
&$ \frac{1}{12}\cdot 2\pi $ &0\\[0.3em]
$\ominus$ & $~+~-~+~$ & $\frac{1}{12}$
&$0\cdot 2\pi $ &4\\[0.3em]
$\ominus$ & $~+~+~-~$ & $\frac{4}{12}$
&$-\frac{3}{12}\cdot 2\pi $ &0\\[0.3em]
\hline
\end{tabular}
\end{center}
\caption{The multiplicity for $s^2=1$ and $(P+6V)\cdot V=\frac16$.
$+$ and $-$ denote $+\frac12$ and $-\frac12$, respectively, and
$\oplus(\ominus)$ is R(L)--handed. Note that $(P+\tilde V)\cdot
V-\frac12 k(V^2-\phi_s^2)=\frac13$.}\label{tbmult}
\end{table}
The components of the vector $\tilde s$ are the last three
$\pm\frac12$'s of $s$.  The number of massless states are given by
$P_{\theta^6}$. For these, we use the Euler numbers
$\tilde\chi(\theta^k,\theta^l)$ given in Table \ref{tb:degn}
\cite{ChoiKim05}, Therefore, we obtain
\begin{equation}
{\cal P}_6={\cal P}_{\theta^6}=\textstyle \frac{1}{12}
\left\{(1+\Delta^6)(16+4\Delta^3)+
\Delta(1+\Delta)(1+\Delta^3+\Delta^6+\Delta^9)\right\}
\end{equation}
which becomes 4, 2, 2, and 3 for $\Delta=1,-1,\Delta^3=-1,+1$,
respectively, and 0 for the other cases. We know that
$s\cdot\tilde\phi=s\cdot\tilde\phi+(\rm integer)$. Now consider the
right mover condition. For $s^2=1$, the masslessness condition
$(s+\tilde\phi)^2=\frac{1}{2}$ leads to $\tilde
s=(-\frac12,\pm\frac12,-\frac12)$ where we used the shifted
$\tilde\phi$ of Eq. (\ref{T6vec}). The relevant ones appear in the
third and fifth rows of Table \ref{tbmult}. Among these one set is
the ${\cal CTP}$ conjugates of the other. The $U(1)_X$ charge is
$(P+6V)\cdot Q_X=5$. Thus, we obtain two singlets as shown in the
third row of Table \ref{T6states}.

Consider $P=(-1~0~0~0~0~-1~0~0)(0^8)$ and
$P=(-1~-1~-1~-1~0~0~0~0)(0^8)$. For $P=(-1~0~0~0~0~-1~0~0)(0^8)$,
\begin{equation}
(P+6V)\cdot V=\textstyle \frac{1}{6},\ \ \tilde \phi\cdot
\phi_s=\textstyle \frac14.
\end{equation}
For $k=6,$ the phase becomes $(P+\tilde V)\cdot V-(\tilde s+ \tilde
\phi)\cdot \phi_s-\frac12 k(V^2-\phi_s^2)=(-\frac{7}{12}-\tilde
s\cdot\phi_s).$ We add $-\frac{8}{12}\cdot 2\pi$ to the fourth
column entries of Table \ref{tbmult}. The masslessness condition
chooses $\tilde s=(-\frac12,\pm\frac12,-\frac12)$, the third and
fifth rows again, leading to the multiplicity 2.
$(\underline{-1~0~0~0~0}-1~0~0)$ is $\overline{\bf 5}$ whose
$U(1)_X$ charge is $3$. For $P=(-1~-1~-1~-1~0~0~0~0)(0^8)$,
\begin{equation}
(P+6V)\cdot V=\textstyle -\frac{1}{6},\ \ \tilde \phi\cdot
\phi_s=\textstyle \frac14.
\end{equation}
Now the phase becomes $(P+\tilde V)\cdot V-(\tilde s+ \tilde
\phi)\cdot \phi_s-\frac12 k(V^2-\phi_s^2)=(-\frac{11}{12}-\tilde
s\cdot\phi_s)=(\frac{1}{12}-\tilde s\cdot\phi_s).$  Thus we obtain
the fourth column entries of Table \ref{tbmult}. The masslessness
condition chooses $\tilde s=(-\frac12,\pm\frac12,-\frac12)$, the
third and fifth rows again, leading to the multiplicity 2.
$(\underline{-1~-1~-1~-1~0}~0~0~0)$ is ${\bf 5}$ whose $U(1)_X$
charge is $-3$. The above four cases are shuffled to list the ${\cal
CTP}$ conjugates together,
\begin{align}
&{\bf 5}_{3,R}{\Big\{ }\begin{split}
 {\bf 5}:\
&(P+6V)=\textstyle(\underline{-\frac12~-\frac12~-\frac12~-\frac12~\frac12}~
\frac12~0~0),\oplus,\quad Q_X=3\\
\overline{\bf 5}:\
&(P+6V)=\textstyle(\underline{-\frac12~\frac12~\frac12~\frac12~\frac12}~
-\frac12~0~0),\ominus,\quad Q_X=-3\\
\end{split}\\
&\overline{\bf 5}_{-3,R}{\Big\{ }\begin{split} \overline{\bf 5}:\
&(P+6V)=\textstyle(\underline{-\frac12
~\frac12~\frac12~\frac12~\frac12}~
-\frac12~0~0),\oplus,\quad Q_X=-3\\
{\bf 5}:\
&(P+6V)=\textstyle(\underline{-\frac12~-\frac12~-\frac12~-\frac12~\frac12}~
\frac12~0~0),\ominus,\quad Q_X=3
\end{split}
\end{align}
Thus we obtain the first and fourth row entries of Table
\ref{T6states}.

Next consider $\overline{\bf 10}$s and ${\bf 10}$s. So consider
$P=(-1~-1~0~0~0~0~0~0)(0^8)$ and $P=(-1~-1~-1~0~0~-1~0~0)(0^8)$. For
$P=(-1~-1~0~0~0~0~0~0)(0^8)$, we have
\begin{equation}
(P+6V)\cdot V=\textstyle \frac{1}{3},\ \ \tilde \phi\cdot
\phi_s=\textstyle \frac14.
\end{equation}
The phase becomes $(P+\tilde V)\cdot V-(\tilde s+ \tilde \phi)\cdot
\phi_s-\frac12 k(V^2-\phi_s^2)=(-\frac{5}{12}-\tilde
s\cdot\phi_s)=(\frac{7}{12}-\tilde s\cdot\phi_s).$  We add
$\frac{6}{12}\cdot 2\pi$ to the fourth column entries of Table
\ref{tbmult}. The masslessness condition for right movers chooses
$\tilde s=(-\frac12,\pm\frac12,-\frac12)$, the third and fifth rows
again, leading to the multiplicity 3 and 4, respectively. For
$P=(-1~-1~-1~0~0~-1~0~0)(0^8)$, we have
\begin{equation}
(P+6V)\cdot V=\textstyle -\frac{1}{3},\ \ \tilde \phi\cdot
\phi_s=\textstyle \frac14.
\end{equation}
The phase becomes $(P+\tilde V)\cdot V-(\tilde s+ \tilde \phi)\cdot
\phi_s-\frac12 k(V^2-\phi_s^2)=(-\frac{1}{12}-\tilde
s\cdot\phi_s)=(\frac{11}{12}-\tilde s\cdot\phi_s).$  We add
$-\frac{2}{12}\cdot 2\pi$ to the fourth column entries of Table
\ref{tbmult}. The masslessness condition chooses $\tilde
s=(-\frac12,\pm\frac12,-\frac12)$, the third and fif rows again,
leading to the multiplicity 4 and 3, respectively.  These four cases
are shuffled to list the ${\cal CTP}$ conjugates together,
\begin{align}
&\overline{\bf 10}_{\bf -1}^{R}{\Big\{ }\begin{split}
 \overline{\bf 10}:\
&(P+6V)=\textstyle(\underline{-\frac12~-\frac12~\frac12~\frac12~\frac12}~
\frac12~0~0),\oplus,{\cal P}_6=3,\quad Q_X=-1\\
{\bf 10}:\
&(P+6V)=\textstyle(\underline{-\frac12~-\frac12~-\frac12~\frac12~\frac12}~
-\frac12~0~0),\ominus,{\cal P}_6=3,\quad Q_X=1\\
\end{split}\\
&{\bf 10}_{1}^{R}{\Big\{ }\begin{split} {\bf 10}:\
&(P+6V)=\textstyle(\underline{-\frac12
~-\frac12~-\frac12~\frac12~\frac12}~
-\frac12~0~0),\oplus,{\cal P}_6=4,\quad Q_X=1\\
\overline{\bf 10}:\
&(P+6V)=\textstyle(\underline{-\frac12~-\frac12~\frac12~\frac12~\frac12}~
\frac12~0~0),\ominus,{\cal P}_6=4,\quad Q_X=-1
\end{split}
\end{align}
Thus we obtain the second and fifth row entries of Table
\ref{T6states}.

Other singlets with nonvanishing oscillators are also allowed, which
are shown in Table \ref{T6states}.

\subsection{Twisted sector $T_1$}

The massless condition for the right mover in the $T_1$ sector is
$(s+\phi_s)^2=\frac{11}{24}$. It allows only one right-handed state
$s=(\ominus ---)$, which gives $(\tilde{s}+\phi_s)\cdot
\phi_s=-\frac{1}{8}$.

Since $-\frac{k}{2}(V^2-\phi^2)=-\frac{1}{4}$ for $k=1$, the phase
in Eq.~(\ref{phase0}) is given by $\Theta_0=(P+V)\cdot
V+\sum_j(N^L)_j(\hat{\phi})_j-\frac{1}{8}$.

The $T_1$ sector is distinguished by Wilson lines: $V_0=V,
V_+=V+a_1,$ and $V_-=V-a_1$. These sectors are denoted as
$T_{1}^0,T_{1}^+$ and $T_{1}^-$.

Note the degeneracy factor $\tilde\chi(\theta^k,\theta^l)$ for
$k=1,2,5$ in Table \ref{tb:degn}. They take the same value 3 along
each horizontal line, as in the prime orbifolds such as $\Z_3$ and
$\Z_7$. Thus, only the states with vanishing phase turn out to
survive the projection by Eq.~(\ref{projP}) in the $T_1$, $T_2$,
$T_5$ (and also in $T_{11}$, $T_{10}$, $T_{7}$) sectors.

The masslessness condition for the left movers (\ref{condit3}) gives
\begin{equation}
{(P+kV)^2}=-{2}\sum_jN^L_j\tilde{\phi}_j+{2}\tilde c.
\end{equation}
For $k=1$ we have $2\tilde c=\frac{35}{24}$. The states satisfying
the massless condition and $(P+V_0)\cdot
V_0+\sum_j(N^L)_j(\hat{\phi})_j=\frac{1}{8}$ for $T_{1}^0$ are
listed in Table \ref{tb:T10}. The multiplicity 3 reduces to 1 due
to the distinction by Wilson lines: $V_0, V_+,$ and $V_-$.

\begin{table}
\begin{center}
\begin{tabular}{|c|c|c|c|}
\hline  $P+V$ & $(N^L)_j$ & ${\cal P}_1(f_0)$ &$\chi$
\\
\hline
 $\left(\overline{{\bf 5}}_{\bf -1/2};\frac{-7}{12},\frac{6}{12},0\right)
 (\frac{1}{6},\frac{1}{6};0^6)'$ & $0$ & $1$ & $L$
 \\
$\left(\overline{{\bf 5}}_{\bf
-1/2};\frac{5}{12},\frac{-6}{12},0\right)
 (\frac{1}{6},\frac{1}{6};0^6)'$ & $1_3$ & $1$ & $L$
 \\
  $\left({\bf 5}_{\bf 1/2};\frac{-1}{12},0,\frac{6}{12}\right)
  (\frac{1}{6},\frac{1}{6};0^6)'$ & $2_3$ & $1$ & $L$
 \\
  $\left({\bf 1}_{\bf -5/2};\frac{-7}{12},\frac{-6}{12},0\right)
  (\frac{1}{6},\frac{1}{6};0^6)'$ & $3_3$ & $1$ & $L$
 \\
$\left({\bf 1}_{\bf -5/2};\frac{5}{12},\frac{6}{12},0\right)
(\frac{1}{6},\frac{1}{6};0^6)'$ & $1_2,~4_3$ & $1+1$ &$L$
 \\
  $\left({\bf 1}_{\bf 5/2};\frac{11}{12},0,\frac{6}{12}\right)
  (\frac{1}{6},\frac{1}{6};0^6)'$ & $0$ & $1$ & $L$
 \\
 $\left({\bf 1}_{\bf 5/2};\frac{-1}{12},0,\frac{-6}{12}\right)
(\frac{1}{6},\frac{1}{6};0^6)'$
& $1_{1},~5_3,~\{1_2+1_3\}$ & $1+1+1$ &$L$
\\
\hline \hline $P+V_+$ & $(N^L)_j$ & ${\cal P}_1(f_+)$&$\chi$
\\
\hline  $\left({\bf 1}_{\bf
-5/2};\frac{5}{12},\frac{2}{12},\frac{4}{12}\right)
(\underline{\frac{-5}{6},\frac{1}{6}};\frac{-1}{3};0^5)'$ & $0$ &
$1$ &$L$
\\
 $\left({\bf
1}_{\bf -5/2};\frac{5}{12},\frac{2}{12},\frac{4}{12}\right)
(\frac{1}{6},\frac{1}{6};\frac{2}{3};0^5)'$ & $2_3$ & $1$ &$L$
\\
 $\left({\bf 1}_{\bf 5/2};\frac{-1}{12},
\frac{-4}{12},\frac{-2}{12}\right)
(\underline{\frac{-5}{6},\frac{1}{6}};\frac{-1}{3};0^5)'$ & $1_3$ &
$1$ &$L$
\\
 $\left({\bf
1}_{\bf 5/2};\frac{-1}{12},\frac{-4}{12},\frac{-2}{12}\right)
(\frac{1}{6},\frac{1}{6};\frac{2}{3};0^5)'$ & $3_3$ & $1$ &$L$
\\
\hline \hline $P+V_-$ & $(N^L)_j$ & ${\cal P}_1(f_-)$&$\chi$
\\
\hline  $\left({\bf 5}_{\bf
1/2};\frac{-1}{12},\frac{4}{12},\frac{2}{12}\right)
(\frac{1}{6},\frac{1}{6};\frac{-2}{3};0^5)'$ & $0$ & $1$ &$L$
\\
 $\left({\bf
1}_{\bf -5/2};\frac{5}{12},\frac{-2}{12},\frac{8}{12}\right)
(\frac{1}{6},\frac{1}{6};\frac{-2}{3};0^5)'$ & $0$ & $1$ &$L$
\\
 $\left({\bf
1}_{\bf -5/2};\frac{-7}{12},\frac{-2}{12},\frac{-4}{12}\right)
(\frac{1}{6},\frac{1}{6};\frac{-2}{3};0^5)'$ & $1_3$ & $1$ &$L$
\\
 $\left({\bf
1}_{\bf 5/2};\frac{-1}{12},\frac{-8}{12},\frac{2}{12}\right)
(\frac{1}{6},\frac{1}{6};\frac{-2}{3};0^5)'$ & $1_3$ & $1$ &$L$
\\
\hline
\end{tabular}
\end{center}
\caption{Chiral matter fields satisfying $\Theta_{0,\pm}=0$ in the
$T_1^0$ and $T_1^\pm$ sectors.  Here, $\overline{\bf
5}_{-1/2}\equiv\left(\underline{\frac{-3}{4},\left[\frac{1}{4}\right]^4}\right)$,
${\bf
5}_{1/2}\equiv\left(\underline{\frac{3}{4},\left[\frac{-1}{4}\right]^4}\right)$,
and ${\bf 1}_{\pm 5/2}\equiv
(\pm\frac{1}{4},\pm\frac{1}{4},\pm\frac{1}{4},\pm\frac{1}{4},\pm\frac{1}{4}
)$.}\label{tb:T10}
\end{table}

In the $T_{1}^+$-sector with $V_+=V+a_3$, only the states with
$(P+V_+)\cdot V_++(N^L)_j(\hat{\phi})_j=\frac{7}{24}$ survives the
GSO projection by Eq.~(\ref{projP}), which are listed in the
middle part of Table \ref{tb:T10}.

In the $T_{1}^-$-sector with $V_-=V-a_3$, only the states with
$(P+V_-)\cdot V_-+(N^L)_j(\hat{\phi})_j=\frac{5}{8}$ survives the
GSO projection by Eq.~(\ref{projP}), which are listed in the lower
part of Table \ref{tb:T10}.

\subsection{Twisted sector $T_2$}
The massless condition for the right mover in the $T_2$ sector is
$(s+2\phi_s)^2=\frac{1}{2}$. It allows only one right-handed state
$s=(\ominus ---)$, which gives $(\tilde{s}+2\phi_s)\cdot
\phi_s=\frac{1}{6}=-\frac56$.

Since $-\frac{k}{2}(V^2-\phi^2)=-\frac{1}{2}$ for $k=2$, the phase
of $\Delta$ in Eq.~(\ref{phase0}) is given by $\Theta_0=(P+2V)\cdot
V+\sum_j(N^L)_j(\hat{\phi})_j+\frac{1}{3}$. The $T_2$ sector is also
distinguished by Wilson lines: $V_0=V, V_+=V+a_3,$ and $V_-=V-a_3$.
These sectors are denoted as $T_{2}^0,T_{2}^+$ and $T_{2}^-$.

Since $\tilde{\chi}(\theta^k,\theta^l)$ with $k=2$ in
Eq.~(\ref{projP}) takes the same value 3 along the horizontal line,
only the states with $\Theta_{0,\pm}=0$ survive the projection
operator in the $T_2$ sector. In the $T_2^0$ sector, hence, the
massless states satisfying the condition $(P+2V)\cdot V+\sum_j
(N^L)_j(\hat{\phi})_j=-\frac{1}{3}$ are selected. Similarly, in the
$T_{2}^{+}$ and $T_{2}^{-}$ sectors, $(P+2V_+)\cdot
V_++\sum_j(N^L)_j(\hat{\phi})_j=0$ and $(P+2V_-)\cdot
V_-+\sum_j(N^L)_j(\hat{\phi})_j=-\frac{1}{3}$ should be chosen. The
massless condition for the left mover is
$|P+2V_{(\pm)}|^2+2\sum_j(N^L)_j(\tilde{\phi})_j=\frac{3}{2}$. The
allowed shifted $E_8\times E_8'$ weight vectors $(P+2V)$s in the
$T_2$ sector are shown in Table \ref{tb:T20}.

\begin{table}
\begin{center}
\begin{tabular}{|c|c|c|c|c|}
\hline  $P+2V$ & $(N^L)_j$ & ${\cal P}_2(f_0)$ &$\chi$ & Labels
\\
\hline $\left({\bf{5}}_{\bf
3};\frac{-1}{6},0^2\right)(\frac{1}{3},\frac{1}{3};0^6)'$ & $0$ &
$1$ &$L$ &
\\
$\left({\bf 1}_{\bf
-5};\frac{-1}{6},0^2\right)(\frac{1}{3},\frac{1}{3};0^6)'$ & $0$ &
$1$ &$L$ &
\\
\hline $\left({\bf 1}_{\bf
0};\frac{1}{3},\frac{1}{2},\frac{-1}{2}\right)(\frac{1}{3},\frac{1}{3};0^6)'$
& $2_{\bar{1}},~2_3$ & $1+1$ &$L$ & $C_1^0,~C_2^0$
 \\
$\left({\bf 1}_{\bf
0};\frac{1}{3},\frac{-1}{2},\frac{1}{2}\right)(\frac{1}{3},\frac{1}{3};0^6)'$
& $1_{\bar{2}},~\{1_{\bar{1}}+1_3\}$ & $1+1$ &$L$ & $C_3^0,~C_4^0$
\\
$\left({\bf 1}_{\bf
0};\frac{-2}{3},\frac{1}{2},\frac{1}{2}\right)(\frac{1}{3},\frac{1}{3};0^6)'$
& $1_{\bar{1}}$ & $1$ &$L$ & $C_5^0$
\\
$\left({\bf 1}_{\bf
0};\frac{-2}{3},\frac{-1}{2},\frac{-1}{2}\right)(\frac{1}{3},\frac{1}{3};0^6)'$
& $1_3$ & $1$ &$L$ & $C_6^0$
\\
$\left({\bf 1}_{\bf
0};\frac{1}{3},\frac{-1}{2},\frac{1}{2}\right)(\frac{-2}{3},\frac{-2}{3};0^6)'$
& $0$ & $1$ &$L$ & $C_7^0$
\\
\hline \hline $P+2V_+$ & $(N^L)_j$ & ${\cal P}_2(f_+)$& $\chi$ &
Labels
\\
\hline $\left({\bf 1}_{\bf
0};\frac{1}{3},\frac{-1}{6},\frac{1}{6}\right)
(\underline{\frac{-2}{3},\frac{1}{3}};\frac{1}{3};0^5)'$ &
$1_{\bar{2}},~\{1_{\bar{1}}+1_3\}$ & $1+1$ & $L$ & $D_1^-,~D_2^-$
\\
$\left({\bf 1}_{\bf
0};\frac{1}{3},\frac{-1}{6},\frac{1}{6}\right)(\frac{1}{3},
\frac{1}{3};\frac{-2}{3};0^5)'$ & $2_{\bar{1}},~ 2_3$ & $1+1$ &
$L$ & $C_1^-,~C_2^-$
\\
\hline \hline $P+2V_-$ & $(N^L)_j$ & ${\cal P}_2(f_-)$& $\chi$ &
Labels
\\
\hline $\left({\bf 1}_{\bf
0};\frac{1}{3},\frac{1}{6},\frac{-1}{6}\right)
(\underline{\frac{-2}{3},\frac{1}{3}};\frac{-1}{3};0^5)'$ &
$2_{\bar{1}},~2_3$ & $1+1$& $L$ & $D_1^+,~D_2^+$
\\
$\left({\bf 1}_{\bf 0};\frac{1}{3},\frac{1}{6},\frac{-1}{6}\right)
(\frac{1}{3},\frac{1}{3};\frac{2}{3};0^5)'$ &
$1_{\bar{2}},~\{1_{\bar{1}}+1_3\}$ & $1+1$& $L$ & $C_1^+,~C_2^+$
\\
\hline
\end{tabular}
\end{center}
\caption{Chiral matter fields satisfying $\Theta_{0}=0,
\Theta_+=0,$ and $\Theta_-=0$ in the $T_2^0, T_2^+$, and $T_2^-$
sectors, respectively. Here ${\bf
1}_0\equiv(0,0,0,0,0)$.}\label{tb:T20}
\end{table}

Consider the first row of Table \ref{tb:T20}. Since
$$
2V=\textstyle
(\frac12~\frac12~\frac12~\frac12~\frac12;~\frac56~1~0)
(\frac13~\frac13~0^6)^\prime,
$$
$P=(\underline{-1~-1~-1~-1~0~}~;~-1~-1~0) (0^8)^\prime$ satisfies
the masslessness condition. The $SU(5)$ representation is $\bf 5$.
Then, we have
\begin{equation}
P+2V=\textstyle
(\underline{-\frac12~-\frac12~-\frac12~-\frac12~+\frac12~}~;~-\frac16~0~0)
(\frac13~\frac13~0^6)^\prime
\end{equation}
which gives $Q_X=3$. From the previous discussion on the right mover
condition, we obtain the chirality $2\ominus$.

For the $T_2^\pm$ sectors, only neutral fields under $\rm
SU(5)\times U(1)_X$ arise, which are tabulated in Table
\ref{tb:T20}.

Thus, from $T_2$ we obtain the following $SU(5)\times U(1)_X$
representations
\begin{eqnarray}
 \{{\bf {5}}_{3},~{\bf 1}_{\bf -5}\}+~ 15\ {\rm neutral~ singlets}.
\end{eqnarray}
The ${\cal CTP}$ {conjugates} are provided from  $T_{10}$.

\subsection{Twisted sector $T_3$}

The shifted momenta in the $T_3$-sector must satisfy $(P+3V)\cdot
a_3=0$ mod. Z, viz. Eq. (\ref{Wilsontw}). It turns out that there is
no massless states satisfying this condition.

\subsection{Twisted sector $T_4$}
The massless condition for the right movers in the $T_4$ sector is
$(s+4\phi_s)^2=\frac{1}{3}$. Taking a shifted $4\phi_s$ as
$(\frac23~\frac13~\frac13)$,  only the right-handed state
$s=(\ominus ---)$ satisfies this condition. So it is left handed.
Now, $(\tilde s+4\phi_s)\cdot\phi_s=0$.

Since $-\frac{k}{2}(V^2-\phi^2)=(\rm integer)$ for $k=4$, it does
not contribute to the phase. So, $\Delta$ of Eq.~(\ref{phase0}) is
given by $\Theta_f=(P+4V_{(\pm)})\cdot
V_{(\pm)}+\sum_j(N^L)_j(\hat{\phi})_j$. The $T_4$ sector is again
distinguished by Wilson lines: $V_0=V, V_+=V+a_3$ and $V_-=V-a_3$.
These sectors are denoted as $T_{4}^0,T_{4}^+$ and $T_{4}^-$.

$\tilde{\chi}(4,n)$ of the $T_4$ sector are $(27, 3, 3, 3)^3$,
hence the multiplicity is
\begin{equation}
P_4=\frac{3}{12}(1+\Delta^4+\Delta^8)(8+[1+\Delta
+\Delta^2+\Delta^3]).
\end{equation}
So, $\Delta=e^{2\pi i/12}, e^{2\pi i/6}$ give ${\cal P}_4=0$. ${\cal
P}_4=9,6,6$ for $\Delta=1,-1,e^{\pm 2\pi i/4}$. Considering Wilson
lines, the nonvanishing multiplicities are $3,2,2$ in each $T_4$.
The massless fields of $T_4^0$ are listed in Table \ref{tb:T4}.

\begin{table}
\begin{center}
\begin{tabular}{|c|c|c|c|c|c|}
\hline  $P+4V$ & $(N^L)_j$ &
 $\Theta_0$ & ${\cal P}_4(f_0)$ &$\chi$ & Labels
\\
\hline $\left({\bf{5}}_{\bf -2};\frac{-1}{3},0^2\right)
(\frac{-1}{3},\frac{-1}{3};0^6)'$ & $0$ & $0$ & $3$ &$L$ &
\\
$\left(\overline{\bf 5}_{\bf 2}
;\frac{-1}{3},0^2\right)(\frac{-1}{3},\frac{-1}{3};0^6)'$ & $0$ &
$\frac{1}{2}$ & $2$&$L$ &
\\
\hline
$$ & $1_{3}$ & $\frac{1}{4}$ & $2$&$L$ & $s_1^0$
\\
$\left({\bf 1}_{\bf
0};\frac{2}{3},0^2\right)(\frac{-1}{3},\frac{-1}{3};0^6)'$ &
$1_{2}$ & $\frac{1}{2}$ & $2$ & $L$ & $s_2^0$
\\
$$ & $1_{\bar{1}}$ & $\frac{-1}{4}$ & $2$&$L$ & $s_3^0$
\\
\hline $\left({\bf 1}_{\bf
0};\frac{2}{3},0^2\right)(\frac{2}{3},\frac{2}{3};0^6)'$ & $0$ &
$\frac{1}{2}$ & $2$&$L$ & $s_4^0$
\\
$\left({\bf 1}_{\bf 0};\frac{-1}{3},\pm
1,0\right)(\frac{-1}{3},\frac{-1}{3};0^6)'$ & $0$ & $\frac{1}{4}$
& $2+2$&$L$ & $s_5^0,~s_6^0$
\\
$\left({\bf 1}_{\bf 0};\frac{-1}{3},0,\pm
1\right)(\frac{-1}{3},\frac{-1}{3};0^6)'$ & $0$ & $\frac{-1}{4}$ &
$2+2$&$L$ & $s_7^0,~s_8^0$
\\
\hline\hline $P+4V_+$ & $(N^L)_j$ & $\Theta_+$ & ${\cal P}_4(f_+)$
&$\chi$ & Labels
\\
\hline $\left({\bf 1}_{\bf
0};\frac{2}{3},\frac{-1}{3},\frac{1}{3}\right)
(\underline{\frac{2}{3},\frac{-1}{3}};\frac{-1}{3};0^5)'$ & $0$ &
$\frac{1}{2}$ & $2$&$L$ & $d_1^+$
\\
$\left({\bf 1}_{\bf 0};\frac{2}{3},\frac{-1}{3},\frac{1}{3}\right)
(\frac{-1}{3},\frac{-1}{3};\frac{2}{3};0^5)'$ & $0$ & $0$ &
$3$&$L$ & $s_1^+$
\\
\hline $\left({\bf 1}_{\bf
0};\frac{-1}{3},\frac{2}{3},\frac{1}{3}\right)
(\underline{\frac{2}{3},\frac{-1}{3}};\frac{-1}{3};0^5)'$ & $0$ &
$\frac{1}{4}$ & $2$&$L$ & $d_2^+$
\\
$\left({\bf 1}_{\bf 0};\frac{-1}{3},\frac{2}{3},\frac{1}{3}\right)
(\frac{-1}{3},\frac{-1}{3};\frac{2}{3};0^5)'$ & $0$ &
$\frac{-1}{4}$ & $2$&$L$ & $s_2^+$
\\
\hline $\left({\bf 1}_{\bf
0};\frac{-1}{3},\frac{-1}{3},\frac{-2}{3}\right)
(\underline{\frac{2}{3},\frac{-1}{3}};\frac{-1}{3};0^5)'$ & $0$ &
$\frac{-1}{4}$ & $2$&$L$ & $d_3^+$
\\
$\left({\bf 1}_{\bf
0};\frac{-1}{3},\frac{-1}{3},\frac{-2}{3}\right)
(\frac{-1}{3},\frac{-1}{3};\frac{2}{3};0^5)'$ & $0$ &
$\frac{1}{4}$ & $2$&$L$ & $s_3^+$
\\
\hline \hline  $P+4V_-$ & $(N^L)_j$ & $\Theta_-$ & ${\cal
P}_4(f_-)$&$\chi$ & Labels
\\
\hline $\left({\bf 1}_{\bf
0};\frac{2}{3},\frac{1}{3},\frac{-1}{3}\right)
(\underline{\frac{2}{3},\frac{-1}{3}};\frac{1}{3};0^5)'$ & $0$ &
$\frac{1}{2}$ & $2$&$L$ & $d_1^-$
\\
$\left({\bf 1}_{\bf 0};\frac{2}{3},\frac{1}{3},\frac{-1}{3}\right)
(\frac{-1}{3},\frac{-1}{3};\frac{-2}{3};0^5)'$ & $0$ & $0$ &
$3$&$L$ & $s_1^-$
\\
\hline $\left({\bf 1}_{\bf
0};\frac{-1}{3},\frac{-2}{3},\frac{-1}{3}\right)
(\underline{\frac{2}{3},\frac{-1}{3}};\frac{1}{3};0^5)'$ & $0$ &
$\frac{1}{4}$ & $2$&$L$ & $d_2^-$
\\
$\left({\bf 1}_{\bf
0};\frac{-1}{3},\frac{-2}{3},\frac{-1}{3}\right)
(\frac{-1}{3},\frac{-1}{3};\frac{-2}{3};0^5)'$ & $0$ &
$\frac{-1}{4}$ & $2$&$L$ & $s_2^-$
\\
\hline $\left({\bf 1}_{\bf
0};\frac{-1}{3},\frac{1}{3},\frac{2}{3}\right)
(\underline{\frac{2}{3},\frac{-1}{3}};\frac{1}{3};0^5)'$ & $0$ &
$\frac{-1}{4}$ & $2$&$L$ & $d_3^-$
\\
$\left({\bf 1}_{\bf 0};\frac{-1}{3},\frac{1}{3},\frac{2}{3}\right)
(\frac{-1}{3},\frac{-1}{3};\frac{-2}{3};0^5)'$ & $0$ &
$\frac{1}{4}$ & $2$&$L$ & $s_3^-$
\\
\hline
\end{tabular}
\end{center}
\caption{Chiral matter fields in the $T_4^0, T_4^+$ and $T_4^-$
sectors. }\label{tb:T4}
\end{table}

Consider first two rows of Table \ref{tb:T4}. They are left handed.
The massless condition for the left movers is $(P+4V)^2=\frac43$.
Since
$$
4V=\textstyle
(1~1~1~1~1~;~\frac53~2~0)(\frac23~\frac23~0^6)^\prime,
$$
the state
$$
(\underline{0,[-1]^4};-2,-2,0)(-1,-1;0^6)'
$$
satisfies the masslessness condition. It is $\bf 5$ and $Q_X=-2$
since $P+4V=(\underline{1~0~0~0~0}\cdots)()^\prime$. Since
$(P+4V)\cdot V=0$, we obtain $\Delta=e^{2\pi i\cdot 0}$ from
(\ref{projP}) and the multiplicity is 3. The state $\overline{\bf
5}$ with $Q_X=2$
$$
(\underline{-2,[-1]^4};-2,-2,0)(-1,-1;0^6)'
$$
also satisfies the masslessness condition.  Since $(P+4V)\cdot
V=-\frac{1}{2}$, we obtain $\Delta=e^{2\pi i\cdot(-\frac12)}$ from
(\ref{projP}) and the multiplicity is 2.

In the $T_{4}^{\pm}$-sectors, the phases in Eq.~(\ref{phasef}) are
respectively give by
\begin{eqnarray}\label{phase4+}
&&\Theta_+=(P+4V_+)\cdot V_++\sum_j(N^L)_j(\hat{\phi})_j+\frac{1}{3}
\\ \label{phase4-}
&&\Theta_-=(P+4V_-)\cdot V_-+\sum_j(N^L)_j(\hat{\phi})_j ~,
\end{eqnarray}
where ``$\frac{1}{3}$'' in Eq.~(\ref{phase4+}) comes from
``$-\frac{4}{2}(V_+^2-\phi_s^2)$.''

It turns out that from the $4(V\pm a_3)$ sectors, only neutral
fields under $\rm SU(5)\times U(1)_X$ arise, which are listed in
Table \ref{tb:T4}.

Thus, the massless states in $T_4(+T_8)$ sectors are
\begin{eqnarray}
{\bf {5}}_{\bf -2}+2 \{ {\bf {5}}_{\bf -2},~\overline{\bf 5}_{\bf 2}
\} + 42\ {\rm neutral~singlets} + {\cal CTP~}~{\rm conjugates}.
\end{eqnarray}

\subsection{Twisted sector $T_5$}

In the $T_5$ (and $T_{5}^{\pm}$) sector of the ${\bf Z}_{12-I}$
orbifold, only the right handed chirality states appear as massless
states from the right mover condition.

The massless condition for the right mover in the $T_5$ sector is
$(s+5\phi_s)^2=\frac{11}{24}$. Taking a shifted $5\phi_s$ as
$(\frac{1}{12}~-\frac{4}{12}~\frac{5}{12})$,  only the right movers
$s=(\oplus -+-)$ satisfies this condition. So it is right handed.
Now, $(\tilde s+5\phi_s)\cdot\phi_s=-\frac18$.

Note that $-\frac{k}{2}(V^2-\phi^2)=-\frac14$ for $k=5$. So, the
phase in Eq.~(\ref{phase0}) is given by $\Theta_0=(P+5V)\cdot
V+\sum_j(N^L)_j(\hat{\phi})_j-\frac18$. The $T_5$ sector is again
distinguished by Wilson lines: $V_0=V, V_+=V+a_3,$ and $V_-=V-a_3$.
These sectors are denoted as $T_{5}^0,T_{5}^+$ and $T_{5}^-$. In the
$T_5$ sectors, only the states with $\Theta_{0,\pm}=0$ survive the
GSO projection Eqs.~(\ref{projP}), (\ref{newprojP}) since all
$\tilde{\chi}(5,l)$s are the same.

Only neutral and vector-like pairs of $SU(5)$ singlets arise in
$T_5$. In the $T_{5}^{+}$-sector, only the states with
$(P+5V_+)\cdot V_++\sum_j(N^L)_j(\hat{\phi})_j=-\frac{1}{24}$
survive the GSO projection Eq.~(\ref{projP}). In the $T_5^-$ sector,
only the states with $(P+5V_-)\cdot
V_-+\sum_j(N^L)_j(\hat{\phi})_j=-\frac{3}{8}$ survive. These are
shown in Table \ref{tb:T50}.

\begin{table}
\begin{center}
\begin{tabular}{|c|c|c|c|}
\hline  $P+5V$ & $(N^L)_j$ & ${\cal P}_5(f_0)$ &$\chi$
\\
\hline
 $\left({\bf 5}_{\bf 1/2};\frac{7}{12},0,\frac{6}{12}\right)
 (\frac{-1}{6},\frac{-1}{6};0^6)'$ & $0$ & $1$ & $R$
 \\
$\left({\bf 5}_{\bf 1/2};\frac{-5}{12},0,\frac{-6}{12}\right)
 (\frac{-1}{6},\frac{-1}{6};0^6)'$ & $1_{1}$ & $1$ & $R$
 \\
  $\left(\overline{{\bf 5}}_{\bf -1/2};\frac{1}{12},\frac{6}{12},0\right)
  (\frac{-1}{6},\frac{-1}{6};0^6)'$ & $2_{1}$ & $1$ & $R$
 \\
  $\left({\bf 1}_{\bf 5/2};\frac{7}{12},0,\frac{-6}{12}\right)
  (\frac{-1}{6},\frac{-1}{6};0^6)'$ & $3_{1}$ & $1$ & $R$
 \\
$\left({\bf 1}_{\bf
5/2};\frac{-5}{12},0,\frac{6}{12}\right)(\frac{-1}{6},\frac{-1}{6};0^6)'$
& $1_{\bar{2}},~4_{1}$ & $1+1$ &$R$
 \\
  $\left({\bf 1}_{\bf -5/2};\frac{-11}{12},\frac{6}{12},0\right)
  (\frac{-1}{6},\frac{-1}{6};0^6)'$ & $0$ & $1$ & $R$
 \\
 $\left({\bf
1}_{\bf
-5/2};\frac{1}{12},\frac{-6}{12},0\right)(\frac{-1}{6},\frac{-1}{6};0^6)'$
& $1_{3},~5_{1},~\{1_{1}+1_{\bar{2}}\}$ & $1+1+1$ &$R$
\\
\hline \hline $P+5V_+$ & $(N^L)_j$ & ${\cal P}_5(f_+)$&$\chi$
\\
\hline  $\left({\bf 1}_{\bf
5/2};\frac{-5}{12},\frac{4}{12},\frac{2}{12}\right)
(\underline{\frac{5}{6},\frac{-1}{6}};\frac{1}{3};0^5)'$ & $0$ &
$1$ &$R$
\\
 $\left({\bf
1}_{\bf 5/2};\frac{-5}{12},\frac{4}{12},\frac{2}{12}\right)
(\frac{-1}{6},\frac{-1}{6};\frac{-2}{3};0^5)'$ & $2_{1}$ & $1$
&$R$
\\
 $\left({\bf
1}_{\bf -5/2};\frac{1}{12},\frac{-2}{12},\frac{-4}{12}\right)
(\underline{\frac{5}{6},\frac{-1}{6}};\frac{1}{3};0^5)'$ & $1_{1}$
& $1$ &$R$
\\
 $\left({\bf
1}_{\bf -5/2};\frac{1}{12},\frac{-2}{12},\frac{-4}{12}\right)
(\frac{-1}{6},\frac{-1}{6};\frac{-2}{3};0^5)'$ & $3_{1}$ & $1$
&$R$
\\
\hline \hline $P+5V_-$ & $(N^L)_j$ & ${\cal P}_5(f_-)$&$\chi$
\\
\hline  $\left(\overline{{\bf 5}}_{\bf
-1/2};\frac{1}{12},\frac{2}{12},\frac{4}{12}\right)
(\frac{-1}{6},\frac{-1}{6};\frac{2}{3};0^5)'$ & $0$ & $1$ &$R$
\\
 $\left({\bf
1}_{\bf 5/2};\frac{-5}{12},\frac{8}{12},\frac{-2}{12}\right)
(\frac{-1}{6},\frac{-1}{6};\frac{2}{3};0^5)'$ & $0$ & $1$ &$R$
\\
 $\left({\bf
1}_{\bf 5/2};\frac{7}{12},\frac{-4}{12},\frac{-2}{12}\right)
(\frac{-1}{6},\frac{-1}{6};\frac{2}{3};0^5)'$ & $1_{1}$ & $1$ &$R$
\\
 $\left({\bf
1}_{\bf -5/2};\frac{1}{12},\frac{2}{12},\frac{-8}{12}\right)
(\frac{-1}{6},\frac{-1}{6};\frac{2}{3};0^5)'$ & $1_{1}$ & $1$ &$R$
\\
\hline
\end{tabular}
\end{center}
\caption{Chiral matter fields satisfying $\Theta_{0,\pm}=0$ in the
$T_5^0$ and $T_5^\pm$ sectors.  }\label{tb:T50}
\end{table}

\section{Summary of matter spectra}
Collecting all the flipped-SU(5) model fields, we obtain the
following:
\begin{equation}
U\ :\ ({\bf 1}_{\bf -5}+{\bf {5}}_{\bf 3}+\overline{\bf 10}_{\bf
-1})^L_{U_3},\quad (\overline{\bf 5}_{\bf 2})_{U_2}^{L},\quad({\bf
1}_{\bf -5}+{\bf {5}}_{\bf 3}+\overline{\bf 10}_{\bf
-1})^L_{U_1},\quad ({\bf 1_0})^L_{U_2},\label{Cuntw}
\end{equation}
from the untwisted sector, and
\begin{align}
&T_6\ :\ {\bf 10}_{\bf 1}^R+\big\{2({\bf 1}_{\bf 5}+{\bf 1}_{\bf
-5}+\overline{\bf 5}_{\bf -3}+ {\bf 5}_{\bf 3})+
3({\bf 10}_{\bf 1}+ \overline{\bf 10}_{\bf -1})\big\}^R+22\{{\bf 1}_{\bf 0}
\}\label{CT6}\\
&T_2\ :\ {\bf 1}_{\bf -5}^L+{\bf 5}_{\bf 3}^L
+15\{{\bf 1}_{\bf 0}\},\label{CT2}\\
&T_4\ :\ {\bf {5}}_{\bf -2}^L+ 2({\bf {5}}_{\bf -2}+\overline{\bf
5}_{\bf 2})^L+42\{{\bf 1}_{\bf 0}\} \label{CT4}
\end{align}
from the twisted sectors. The chiral matter resulting from this
spectra constitutes three families of quarks and leptons.

In addition, we obtain G-exotics and E-exotics from $T_1$ and $T_5$
sectors,
\begin{align}
T_1\ :&\ 2(\overline{\bf 5}_{\bf -\frac12})^L+2({\bf 5}_{\bf
+\frac12})^L +
7({\bf 1}_{\bf -\frac52})^L+7({\bf 1}_{\bf +\frac52})^L,\label{ECT10}\\
T_5\ :&\ 2({\bf 5}_{\bf +\frac12})^R+2(\overline{\bf 5}_{\bf
-\frac12})^R+7({\bf 1}_{\bf +\frac52})^R+ 7({\bf 1}_{\bf
-\frac52})^R, \label{ECT50}
\end{align}
From (\ref{Yflipped}), we note that G-exotics carry
\Qem=$\pm\frac16$ quarks and \Qem=$\pm\frac12$ E-exotics and
E-exotics with $X=\pm\frac52$ have  \Qem=$\pm\frac12$. All these
exotics  are removed if all $U(1)$s except the $U(1)_Y$ are broken
at the GUT scale, and there are not left with dangerous half-integer
charged fields below the GUT scale. The lightest of these
half-integer charged fields(LHIC) is absolutely stable since all the
light observable  SM fields (including color singlet composites) are
integer charged. If the mass of LHIC is much larger than the
reheating temperature after inflation, we expect that most of LHIC
are diluted away by inflation.

For the hidden sector, there appear 20 SU(2)$^\prime$ doublets but
no matter for SO(10)$^\prime$.


\section{Yukawa couplings}

Nonvanishing couplings of vertex operators are constructed by
satisfying the ${\bf Z}_{12-I}$ symmetry \cite{HVyuk}. It is
summarized in \cite{ChoiKim05}.  In our notation for the shift
vector, basically it amounts for the operator ${\cal O}_A{\cal
O}_B{\cal O}_C\cdots$ to satisfy
\begin{align}
&{\rm Invariance\ under\ the\ group\ space\ shift\ } V, {\rm and}\\
&{\rm Invariance\ under\ the\ internal\ space\ shift\ } \phi_s.
\end{align}
For the shift $V$, it is easy to check the modular invariance: The
relevant vertex operators have only to satisfy the gauge
invariance. The invariance under the shift $\phi_s$ belongs to a
generalized Lorentz shift and the condition is sometimes called
the $H$-momentum conservation. The (bosonic) $H$-momentum is
defined as
\begin{eqnarray}
R_i =(\tilde{s}+k\phi_s+\tilde{r})_i - (N^L_i-N^L_{\bar{i}}) ~,
~~~i=\{1,2,3\} ~,
\end{eqnarray}
where $\tilde{r}$ is $(\frac{-1}{2},\frac{1}{2},\frac{1}{2})$ for
left handed states ($\equiv \tilde{r}_-$) and
$(\frac{1}{2},\frac{-1}{2},\frac{-1}{2})$ for right handed states
($\equiv \tilde{r}_+$). As discussed earlier, $\tilde{s}$ should
satisfy the mass shell condition,
$|(\frac{1}{2}\chi,\tilde{s})+k\phi_s|^2=2c$.
%
%
$R_i$ can be interpreted as a discrete $R$-charge. Thus,
neglecting oscillator numbers, the $H$-momenta for ${\bf
Z}_{12-I}$ twist are
\begin{equation}
%
\begin{split}
&U_1: \textstyle (-1~0~0),\quad  U_2: \textstyle (0~1~0),\quad
 U_3: \textstyle (0~0~1),\quad \\
&T_1: \textstyle (\frac{-7}{12}~\frac{4}{12}~\frac{1}{12}),\quad
T_2: (\frac{-1}{6}~\frac{4}{6}~\frac{1}{6}),\quad
 T_3: (\frac14~0~\frac{-3}{4})\\
&T_4:  \textstyle (\frac{-1}{3}~\frac13~\frac13),\quad T_5:
(\frac{1}{12}~\frac{-4}{12}~\frac{-7}{12}),\quad
 T_6:(\frac{-1}{2}~0~\frac12)
 \end{split}
%
 \label{Hmomenta}
\end{equation}
which are used to check the generalized Lorentz invariance.
%
%
%

As an example, consider the $T_6$ $H$-momentum,
$(\frac{-1}{2}~0~\frac12)$. It is derived in the following way.
The right mover mass shell condition is
\begin{equation}
\textstyle\frac12 M_R^2=(s+6\phi_s)^2=2c=\textstyle\frac12
\label{MassT6R}
\end{equation}
where $6\phi_s=(\frac{30}{12}~\frac{24}{12}~\frac{6}{12})$. There
are two solutions of Eq. (\ref{MassT6R}),
$$
s_+=\textstyle(\oplus;\frac{-5}{2},\frac{-3}{2},
\frac{-1}{2}),\quad {\rm and}\
s_-=(\ominus;\frac{-5}{2}~\frac{-5}{2}~\frac{-1}{2}).
$$
For the left handed states appearing in Table~\ref{T6states}, let us
focus on $s_-$ whose corresponding sum-to-odd-integer solution is
$(-3~-2~0)$. So bosonization of $\tilde s_-$ is $\tilde s_-+\tilde
r_-$. Thus, $H$-momentum, in analogy with $P+kV$, is
$(\tilde{s}_-+\tilde{r}_-)+k\phi_s=(\frac{-1}{2}~0~\frac12)$
appearing in Eq. (\ref{Hmomenta}).

The $H$-momenta conservation conditions with
$\phi_s=(\frac{5}{12},\frac{4}{12},\frac{1}{12})$ can be
reformulated only in terms of the bosonic $H$-momenta  as follows:
\begin{eqnarray} \label{select-1}
\sum_z R_1(z)=-1~~{\rm mod.}~12~,~~\sum_zR_2(z)=1~~{\rm
mod.}~3~,~~\sum_zR_3(z)=1~~{\rm mod.}~12~,~~
\end{eqnarray}
where $z$ ($\equiv A,B,C,\cdots$) denotes the index of states
participating in a vertex operator. In addition, space group
selection rules requires a vertex operator with $z$-states in
$T_k^{m_f}$ sector ($k=0$ for the untwisted sector) should satisfy
\begin{eqnarray} \label{select-2}
&&\sum_z k (z) = 0~~~~{\rm mod.}~12~,~~
\\ \label{select-3}
&&\sum_z [k m_f](z) = 0~~{\rm mod.}~3~.
\end{eqnarray}
One can easily check the following cubic couplings are allowed in
$\Z_{12-I}$ orbifold models fulfilling Eqs.~(\ref{select-1}) and
(\ref{select-2}), if $N^L_i=N^L_{\bar{i}}=0$~\cite{HVyuk,YukKob}:
\begin{equation}
 U_1U_2U_3~,~~ T_6T_6U_2~,~~ T_4T_4T_4~,~~ T_2T_4T_6~,~~ T_1T_4T_7~.
 \label{Yukawa}
\end{equation}
Note that in considering the superpotential couplings, one should
consider only the same chirality, and in our model there is no
massless states from the $T_3$ and $T_9$ sectors.

For future convenience, we display the $H$-momenta for
combinations of  some singlets appearing in our model in Table
\ref{listH-mom}.

\begin{table}
\begin{center}
\begin{tabular}{|c|c||c|c|}
\hline Comb. of singlets & $H$-momenta & Comb. of singlets &
$H$-momenta
\\ \hline
$h_1\bar{h}_1$ & $(-1~ 0 -1)$ & $h_1\bar{h}_2$ & $(-1 ~0 ~1)$
\\
$h_2\bar{h}_2$ & $(-1~ 0~ 3)$ & $h_1\bar{h}_4$ & $(-2 ~0 ~0)$
\\
$h_3\bar{h}_3$ & $(1~ 0~ 1)$ & $h_2\bar{h}_3$ & $(0~ 0~ 2)$
\\
$h_4\bar{h}_4$ & $(-3~ 0~ 1)$ & $h_2\bar{h}_4$ & $(-2~ 0~ 2)$
\\
\hline \hline Comb. of doublets & $H$-momenta & Comb. of doublets
& $H$-momenta
\\ \hline
$D_1^-d_2^+\bar{h}_1$ & $(-1~3~0)$ & $D_1^-d_2^+\bar{h}_3$ &
$(0~3~1)$
\\
$D_1^-d_2^+\bar{h}_2$ & $(-1~3~2)$ & $D_1^-d_2^+\bar{h}_4$ &
$(-2~3~1)$
\\
\hline \hline Comb. of singlets & $H$-momenta & Comb. of singlets
& $H$-momenta
\\ \hline
$C_1^-s_2^+\bar{h}_1$ & $(1~1~0)$ & $C_1^+s_2^-h_1$ & $(-1~3~0)$
\\
$C_1^-s_2^+\bar{h}_2$ & $(1~1~2)$ & $C_1^+s_2^-h_2$ & $(-1~3~2)$
\\
$C_1^-s_2^+\bar{h}_3$ & $(2~1~1)$ & $C_1^+s_2^-h_3$ & $(0~3~1)$
\\
$C_1^-s_2^+\bar{h}_4$ & $(0~1~1)$ & $C_1^+s_2^-h_4$ & $(-2~3~1)$
\\
\hline
$C_2^-s_2^+\bar{h}_1$ & $(-1~1-2)$ & $C_2^+s_2^-h_1$ & $(0~1-1)$
\\
$C_2^-s_2^+\bar{h}_2$ & $(-1~1~0)$ & $C_2^+s_2^-h_2$ & $(0~1~1)$
\\
$C_2^-s_2^+\bar{h}_3$ & $(0~1-1)$ & $C_2^+s_2^-h_3$ & $(1~1~0)$
\\
$C_2^-s_2^+\bar{h}_4$ & $(-2~1-1)$ & $C_2^+s_2^-h_4$ & $(-1~1~0)$
\\
\hline
\end{tabular}
\end{center}
\caption{$H$-momenta for some combinations of neutral singlets under
$SU(5)\times U(1)_X$ appearing in our model. All the combinations
are neutral under all gauge symmetries in this model, and fulfill
the space group selection rules. }\label{listH-mom}
\end{table}

\subsection{Flipped SU(5) spectrum}\label{subsec:MSSM}

There exist $({\bf 10}_{1}, \overline{\bf 10}_{\bf -1})$ whose VEV
(in the $SU(2)$ singlet direction $\nu^c$) breaks the flipped
$SU(5)$ down to the standard model. Also, there exist the needed
electroweak Higgs fields $(\overline{\bf {5}}_{2},~{\bf 5}_{\bf
-2})$.

In $T_1$ and $T_5$, there appear G-exotics and E-exotics. These
are removed by $T_1T_4T_7$ couplings of Eq. (\ref{Yukawa}) via
$\langle {\bf 1_0}(T_4)\rangle$ and other singlet's VEVs. For
instance, $({\bf
1_{5/2}};\frac{-1}{12},0,\frac{-6}{12})(\frac{1}{6}\frac{1}{6};0^6)'$
with $N^L_j=1_1$ in $T_1^0$, and $({\bf
1_{-5/2}};\frac{-7}{12},\frac{4}{12},\frac{2}{12})
(\frac{-1}{6}\frac{-1}{6};\frac{2}{3};0^5)'$ with $N^L_j=1_{\bar
1}$ in $T_7^-$ get a mass from the coupling with $\langle
s_1^+\rangle$ in $T_4^+$.
Similarly $({\bf
1_{-5/2}};\frac{-7}{12},\frac{-2}{12},\frac{-4}{12})
(\frac{1}{6}\frac{1}{6};\frac{-2}{3};0^5)'$ with $N^L_j=1_3$ in
$T_1^-$, and $({\bf
1_{5/2}};\frac{-1}{12},\frac{6}{12},0)(\frac{1}{6}\frac{1}{6};0^6)'$
with $N^L_j=1_{\bar{3}}$ in $T_7^0$ achieve a mass also from
$s_1^+$.
It turns out that the contributions of oscillator numbers carried
by the other states to the $H$-momenta can be always cancelled by
(multi-) products of $C_1^-s_3^+h_3$, $C_2^-s_2^+\bar{h}_3$,
$C_1^+s_2^-h_3$, $C_1^+s_2^-\bar{h}_1$, etc. When
$N^L_j=\{1_2+1_3\}$ is involved in a $T_1T_7T_4$ vertex operator,
for instance, one could additionally multiply
$C_1^-s_2^+\bar{h}_4$ to the vertex operator in order to cancel
the oscillator number contribution. When $N^l_j=4_3$ is involved,
e.g. $(C_1^+s_2^-h_3)^4$ needs to be multiplied.

In $T_6$ and $T_4$, there appear vectorlike representations $({\bf
10}_{\bf 1}+\overline{\bf 10}_{\bf -1})$s, $({\bf 1}_{\bf 5}+{\bf
1}_{\bf -5})$s, and $({\bf 5}_{\bf -2}+\overline{\bf {5}}_{\bf
2})$s.
If these are removed by the survival hypothesis, we obtain just only
one ${\bf 5}_{\bf -2}^L$ from $T_4$ and $\overline{\bf{5}}_{\bf
2}^L$ from the untwisted sector. One can anticipate that this may
happen if the twisted sectors provide large Yukawa couplings.
Indeed, such couplings are present.

The fields $2({\bf {5}}_{\bf -2}+\overline{\bf 5}_{\bf 2})^L$ of
$T_4$ can be removed by $T_4T_4T_4$ couplings where one $T_4$ is
$\langle s_4^0\rangle$ in Table~\ref{tb:T4}.
The vectorlike representations in $T_6$ are also removed  by the
couplings, e.g. with $(C_1^-s_2^+\bar{h}_1)(C_1^+s_2^-h_1)$ and
$(C_2^-s_2^+\bar{h}_3)(C_1^+s_2^-h_3)$, etc. Thus, $\big\{2({\bf
1}_{\bf 5}+{\bf 1}_{\bf -5}+\overline{\bf 5}_{\bf -3}+ {\bf 5}_{\bf
3})$ + $3({\bf 10}_{\bf 1}+ \overline{\bf 10}_{\bf -1})\big\}^L$ and
$2({\bf {5}}_{\bf -2}+\overline{\bf 5}_{\bf 2})^L$ are expected to
be removed at the GUT scale. In all these, several singlets with
$Q_X=0$ are expected to develop GUT scale VEVs. Then, there result
the following light fields,
\begin{align}
&U\ :\ ({\bf 1}_{\bf 5}+\overline{\bf {5}}_{\bf -3}+{\bf 10}_{\bf
1})^R_{U_1}+({\bf 1}_{\bf 5}+\overline{\bf {5}}_{\bf -3}+{\bf
10}_{\bf 1})^R_{U_3}+({\bf
5}_{\bf -2})^{R}_{U_2},\label{untmatter}\\
&T_6\ :\ {\bf 10}_{\bf 1}^R,\label{matt6}\\
&T_2^0\ :\ {\bf 1}_{\bf 5}^R+\overline{\bf 5}_{\bf -3}^R,\label{matt2}\\
&T_4\ :\ \overline{\bf {5}}_{\bf 2}^R.\label{matt4}
\end{align}
These constitute the three families and one pair of Higgs quintets.
Relabeling R to L, we can obtain the standard form of left-handed
$W^\pm$ interactions. In the remainder, however, we will keep R to
compare with the entries of tables.

It is interesting to note that the pair of Higgs quintets, $({\bf
5}_{\bf -2}+\overline{\bf {5}}_{\bf 2})$, survives the above
analysis. Certainly, it is not allowed to write $M_{\rm GUT}{\bf
5}_{\bf -2}\overline{\bf {5}}_{\bf 2}$ since there is no coupling of
the form $U_2T_4$. In fact, it is not easy to construct a ${\bf
5}_{\bf -2}\overline{\bf {5}}_{\bf 2}$ consistent with the
$H$-momentum conservation.
Thus Higgs doublet mass can be far below the GUT scale, which is a
good thing. But  the colored scalars in $({\bf 5}_{\bf
-2}+\overline{\bf {5}}_{\bf 2})$ must be removed at high energy
scale toward MSSM. It is achieved by the doublet-triplet splitting
mechanism we discuss below.

\subsection{GUT breaking}

The GUT breaking in the flipped-SU(5) model proceeds by $\langle{\bf
10}_{\bf 1}\rangle=\langle\overline{\bf 10}_{\bf -1}\rangle=M_{\rm
GUT}$, which is a D-flat direction. This vector-like representation,
${\bf 10}_{\bf 1}+\overline{\bf 10}_{\bf -1}$, is present in the
$T_6$ sector.  Note that the $H$-momentum of $({\bf 10}_{\bf
1}\overline{\bf 10}_{\bf -1})^L$ is $(-1,0,1)$.

The $T_6$ sector also contains twenty-two $Q_{\rm em}=0$ singlets.
Some combinations of them e.g. $h_1\bar{h}_2$ in
Table~\ref{listH-mom} also give the $H$-momentum of $(-1,0,1)$.
$h_3\bar{h}_4$, $h_2\bar{h}_1$, $h_4\bar{h}_3$ also provide the
same $H$-momentum.  The GUT scale VEV of them could induce
$\langle{\bf 10}_{\bf 1}\rangle=\langle\overline{\bf 10}_{\bf
-1}\rangle=M_{\rm GUT}$ along an F-flat direction, for instance,
through a non-renormalizeable superpotential,
\begin{eqnarray} \label{GUTbreaking}
W=(D_1^- d_2^+ \bar{h}_1 )(C_1^- s_2^+\bar{h}_1)\bigg[{\bf
10}^H_{\bf 1}\overline{\bf 10}^H_{\bf -1}
-h_1\bar{h}_2-h_3\bar{h}_4+\cdots\bigg] ~
\end{eqnarray}
with $\langle D_1^-\rangle$ or $\langle d_2^+\rangle$ vanishing.
Note that there are 20 $\rm SU(2)'$ doublets. Thus, the gauge
coupling of $\rm SU(2)'$ would not blow up at lower energies.
The neutrino direction of $\langle{\bf 10}_{\bf 1}\rangle$,
$\langle\overline{\bf 10}_{\bf -1}\rangle$ allows the symmetry
breaking
\begin{equation}
SU(5)\times U(1)_X\to SU(3)_c\times SU(2)\times U(1)_Y.
\end{equation}

\subsection{Doublet-triplet splitting}

Let us call the three vectorlike ten and tenbars of Eq. (\ref{CT6})
as Higgs fields,  ${\bf 10}_{\bf 1}^H$ and $\overline{\bf 10}_{\bf
-1}^H$. Among the three two are purely vectorlike and removed at the
GUT scale. One remaining vectorlike pair joins the Higgs mechanism.
Let us consider this remaining pair.

The  Higgs ${\bf 10}_{\bf 1}^H$ and $\overline{\bf 10}_{\bf -1}^H$
contain $\{q,~d^c,~\nu^c\}+\{
\overline{q},~\overline{d}^c,~\overline{\nu}^c\}$. in terms of the
standard model quantum numbers. $\{\nu^c,~\overline{\nu}^c\}$ obtain
GUT scale masses from Eq.~(\ref{GUTbreaking}) when $\rm SU(5)\times
U(1)_X$ is broken. $\{q,~\overline{q}\}$ contained in ${\bf 10}_{\bf
1}^H$ and $\overline{\bf 10}_{\bf -1}^H$ are absorbed by the heavy
gauge sector. On the other hand, $\{ d^c,~\overline{d}^c\}$ still
remain light. In order to make the standard model vacuum stabilized,
somehow they should get superheavy masses.

Let us consider $W={\bf 10}_{\bf 1}^H{\bf 10}_{\bf 1}^H{\bf 5}_h
+\overline{\bf 10}_{\bf -1}^H\overline{\bf 10}_{\bf
-1}^H{\overline{\bf 5}}_h$, where ${\bf 5}_h$ and $\overline{\bf
{5}}_h$ indicate the five-plets inducing electroweak symmetry
breaking~\cite{Hwang,Huang:2006nu}. When ${\bf 10}_{\bf 1}^H$ and
$\overline{\bf 10}_{\bf -1}^H$ develop VEVs in  the
$\langle\nu^c\rangle=\langle\overline{\nu}^c\rangle=M_{\rm GUT}$
direction, $\{ d^c,~\overline{d}^c\}$ in ${\bf 10}_{\bf 1}^H$,
$\overline{\bf 10}_{\bf -1}^H$ and triplets ($D,\overline{D}$)
included in ${\bf 5}_h$, $\overline{\bf {5}}_h$ pair up to be
superheavy~\cite{Antoniadis:1987dx}. This achieves the
doublet-triplet splitting in flipped SU(5).

This mechanism can be realized also in our model. The
$[\overline{\bf 10}_{\bf -1}^H\overline{\bf 10}_{\bf
-1}^H\overline{\bf 5}_h]^L $ term arises from $T_6T_6U_2$, which
satisfies the $H$-momentum conservation. The $[{\bf 10}_{\bf
1}^H{\bf 10}_{\bf 1}^H{{\bf 5}}_h]^L$ is still also possible from
highly non-renormalizable interactions, e.g. $[{\bf 10}_{\bf
1}^H{\bf 10}_{\bf 1}^H{{\bf 5}}_h]^L
(s_1^+s_1^-s^u)(C_2^-s_2^+\bar{h}_3)(C_2^+s_2^-h_3)$, which
satisfies all selection rules discussed above. Thus we should assume
cutoff scale VEVs for the neutral singlets.

 Thus, we obtain the so-called MSSM spectra with one
pair of Higgs doublets at cubic level. However, the Higgs doublets
would obtain mass since there exist a lot of singlets. Indeed, there
exist many couplings. The simplest two terms for $\mu$ are
\begin{eqnarray}
\textstyle \bigg[({\bf \overline{5}_{2}};-1,0,0)(0^8)'\bigg]_{U_2}
\bigg[({\bf
5_{-2}};\frac{-1}{3},0,0)(\frac{-1}{3},\frac{-1}{3};0^6)'\bigg]_{T_4^0}
\bigg[s^u(s_2^0)^2+f_{\mu}s_2^0s_4^0\bigg]_{[T_4^0]^2}
\end{eqnarray}
where $f_\mu$ is a relative strength. Before, $s_4^0$ was needed for
charged lepton masses. On the other hand, $s_2^0$ was needed for
mixing of charged leptons. So, we can set $s_2^0$ a free parameter.
Assuming that ${\bf \overline{5}_{2}}$ and ${\bf {5}_{-2}}$ obtain
VEVs, the F-flat direction chooses $s_2^0=-f_\mu s_4^0/2s^u$ so that
the $\mu$ term is of order $-f_\mu^2 s_4^{02}/4s^u$. This can be
linked to the charged lepton masses. For example, if we choose
$s_4^0\sim 10^{-5}$ from electron mass and $s^u\sim O(1)$, we
require $f_\mu=O(10^{-2})$ to obtain a TeV scale $\mu$ term. This
fine tuning is a huge improvement over a fine tuning of $10^{-15}$.
Since we neglected many higher order terms, this is just an
illustration of smoothing the fine tuning problem.
 The
$s_2^0$ mimics the axion multiplet of the $\mu$ solution
\cite{KimNilles}.

\subsection{Fermion masses}

$\bullet$ There is one cubic coupling relevant for $u$-type quarks
and Dirac neutrinos, $T_6T_2T_4$, which is interpreted as the top
quark and (Dirac) tau neutrino Yukawa coupling,
\begin{equation}
t~{\rm quark~+~(Dirac)}~\tau~{\rm neutrino\ coupling}\ :\ [{\bf
10_1}\overline{\bf {5}}_{\bf -3}\overline{\bf {5}}_{\bf 2}]^R.
\end{equation}
Non-renormalizable terms allow the other $u$-type quark (Dirac
neutrino) Yukawa couplings. Most of all, through the couplings with
$C_5^0$, $C_6^0$ in $T_2^0$ and $h_1$, $\bar{h}_3$ in $T_6$,
$[\overline{\bf 10}_{-1}(T_6){\bf{5}_{3}}(U_1){\bf
{5}_{-2}}(T_4^0)]^L$, $[\overline{\bf
10}_{-1}(T_6){\bf{5}_{3}}(U_3){\bf {5}_{-2}}(T_4^0)]^L$,
$[\overline{\bf 10}_{-1}(U_3){\bf{5}_{3}}(T_2^0){\bf
{5}_{-2}}(T_4^0)]^L$, and $[\overline{\bf
10}_{-1}(U_1){\bf{5}_{3}}(T_2^0){\bf {5}_{-2}}(T_4^0)]^L$ are
allowed. With $h_1C_5^0$, $\bar{h}_3C_6^0$, $\bar{h}_1C_5^0$, and
$h_1C_6^0$, the masses and mixing for the first two families of
$u$-type quarks and Dirac neutrinos are possible.

$\bullet$ The bottom quark mass arises in terms of $U_2$  Higgs
doublet. Indeed, such a coupling is present from $T_6T_6U_2$. If
the coupling strength of  $T_6T_2T_4$ and $T_6T_6U_2$ are
comparable, a large $\tan\beta$ is needed to obtain $m_t/m_b\sim
35$. But the couplings depend on the location of the respective
fields and one may treat the ratio as a free parameter of order 1
\cite{kkl1}.

In fact $[{\bf 10_{1}}(U_1){\bf 10_{1}}(U_3){\bf 5_{-2}}(U_2)]^R$
exists. For smallness of it we could assume a proper volume of the 6
dimensional compact space.  As an alternative way, one could
consider $(1,1)$ and $(2,2)$ components of the $d$-quark mass
matrix. They can be induced with $\langle h_1\bar{h}_1\rangle$ and
$\langle h_3\bar{h}_3\rangle$. By $\langle h_1\rangle$ and
$\langle\bar{h}_3\rangle$, the mixing between the first two and the
third families of $d$-type quarks are also permitted.

$\bullet$ For the mass of charged lepton in the twisted sector
$T_2^0$, we need a coupling containing $[{\bf 1_{-5}}(T_2^0){\bf
5}_{\bf 3}(T_2^0)\overline{\bf 5}_{\bf 2}(U_2)]^L$. This coupling is
possible by being supplemented with $(s_1^0s_3^0)(s_4^0s_2^+s_2^-)$.
On the other hand, cubic couplings $[{\bf 1_5}(U_1)\overline{\bf
5}_{\bf -3}(U_3){\bf 5}_{\bf -2}(U_2)]^R$ $[{\bf
1_5}(U_3)\overline{\bf 5}_{\bf -3}(U_1){\bf 5}_{\bf -2}(U_2)]^R$ are
possible. It may be that $\tau$ is placed in the untwisted sector.
In this sense, the leptonic sector does not go parallel to the quark
sector. The $b-\tau$ unification is not achieved in this model.
$(1,1)$ and $(2,2)$ components of the charged lepton mass matrix are
generated via e.g. $\langle h_1\bar{h}_1\rangle$ and $\langle
h_3\bar{h}_3\rangle$. The mixing terms between the first two and the
third families of charged leptons should be mediated by $\langle
h_3s_2^0\rangle$, $\langle \bar{h}_1s_2^0\rangle$, $\langle
h_3s_2^0\rangle$, and $\langle \bar{h}_1s_2^0\rangle$.

$\bullet$ In flipped SU(5), Majorana neutrino masses arise from $
[\langle{\bf 10_1}\rangle \cdot \langle{\bf 10_1}\rangle\cdot
\overline{\bf 10}_{\bf -1} \cdot  \overline{\bf 10}_{\bf -1}]^L$.
The $H$-momentum of this operator is $(-2,0,2)$. Thus this operator
can be induced when supplemented e.g. by the coupling with
$(C_1^-s_2^+\bar{h}_3)(C_2^-s_2^+\bar{h}_1)[(C_1^+s_2^-h_1)(C_2^+s_2^-h_3)]^2$.
The other components of the Majorana neutrino mass matrix should
require additional singlet VEVs such as $h_1$, $\bar{h}_3$,
$(h_1)^2$, and $(\bar{h}_3)^2$. If heavy Majorana neutrino masses
are around $10^{14}$ GeV, thus we obtain the $\nu_\tau$ mass of
order 0.1 eV,
\begin{equation}
m_{\nu_\tau}\simeq \frac{m_{\rm top}^2}{(10^{14}\ \rm GeV)} \sim
0.1\ \rm eV
\end{equation}
for $m_{\rm top}=m_{\rm Dirac\ neutrino}$, which is valid in
flipped SU(5).

\subsection{$R$-parity}

If we consider cubic couplings Eq.~(\ref{Yukawa}), we can define
an $R$-parity in the standard way, $R=-1$ for matter fermions and
$R=+1$ for Higgs bosons.

Firstly, consider the coupling ${\bf 10_1}({U_1}){\bf 10_1}({U_3})
{{\bf 5}_{\bf -2}}({U_2})$. A nontrivial parity can be defined as
\begin{align}
&R=-1\quad {\rm for\ }\ {\bf 10_{1}}(U_1), \ {\bf 10_1}(U_3),\
{\overline{\bf {5}}_{\bf -3}}(U_1),\ {\overline{\bf {5}}_{\bf
-3}}(U_3),\
{\bf 1_{5}}(U_1),\ {\bf 1_{5}}(U_3) \\
&R= +1\quad {\rm for\ }\ {{\bf 5}_{\bf -2}}(U_2)\ .
\end{align}

As discussed above, mixing between the first two families of
matter in the untwisted sector and the third family of matter in
$T_5$ and $T_2^0$ are always possible if VEVs of some neutral
singlets are supposed. Such neutral singlets should also be inert
under all symmetries relevant at low energies.  Otherwise,
symmetries must be broken at low energies by their VEVs. Even with
the mixing terms between untwisted and twisted matter fields, the
$R$-parity relevant in low energies can still be defined by
assigning $R=1$ for the neutral singlets developing VEVs and
\begin{align}
&R=-1\quad {\rm for\ }\ {\bf \overline{10}_{-1}}^L(T_6),\ {\bf
{5}}_{\bf 3}^L(T_2),\ {\bf 1_{-5}}^L(T_2)~.
\end{align}
Then, the allowed Yukawa coupling $T_6T_2T_4$ determines
\begin{equation}
R= +1\quad {\rm for\ }\ {\bf 5}_{\bf -2}(T_4)\ .
\end{equation}
Thus, $R$-parity can survive down to low energies and hence
$R$-parity conservation for proton longevity is fulfilled in the
present model.


\section{Electroweak hypercharge in flipped SU(5)}

\subsection{GUT value of weak mixing angle}

Flipped SU(5) was originally considered as a subgroup of SO(10)
\cite{antisu5}, which can be called SO(10)-flipped SU(5). In this
case, the GUT value for $\sin^2\theta_W$ should be $\frac38$. In
this case, symmetry breaking must proceed via adjoint Higgs field.
The shift vector must take a form
\begin{equation}
V=\left\{ \begin{split}
 &(0~0~0~0~0~x~y~z)(\cdots)^\prime,\ {\rm or}\\
 &\textstyle(\frac12~\frac12~\frac12~\frac12~\frac12~x~y~z)(\cdots)^\prime
 \end{split}\right.\label{SO10flipped}
\end{equation}
so that an SO(10) group is obtained.

String flipped SU(5), for example Eq. (\ref{shiftV}) with five
$\frac14$s, is basically different from SO(10) flipped SU(5) even
though it includes SO(10) flipped SU(5) if the shift vector takes
the form Eq. (\ref{SO10flipped}). If the electroweak $\rm
SU(2)\times U(1)_Y$ is embedded in a simple group, the GUT value of
$\sin^2\theta_W$ is $\frac38$ as calculated from
\begin{equation}
\sin^2\theta_W=\frac{{\rm Tr}T^2_3}{{\rm Tr}Q_{\rm
em}^2}.\label{sin2tSimple}
\end{equation}

In string compactification, $\sin^2\theta_W$ depends on the spectrum
in the untwisted sector \cite{ChoiKim05}. Therefore, if the
untwisted sector spectrum includes ${\bf 16}_{\rm flip}$, then
$\sin^2\theta_W$ is the same as that of SO(10). On the other hand,
if ${\bf 16}_{\rm flip}$ cannot be obtained from the fields in the
$U$ sector, $\sin^2\theta_W=\frac38$ is not warranted. In the
present model, a complete multiplet ${\bf 16}_{\rm flip}$ appears in
the $U$ sector, we obtain $\sin^2\theta_W=\frac38$. This is
explicitly shown below.

The electroweak hypercharge $Y$ is a combination of SU(5) and $\rm
U(1)_X$ generators,  $Y=\frac{1}{5}(X+Y_5)$. As will be shown in
the next section, the $\rm U(1)_Y$ gauge coupling is
\begin{eqnarray} \label{Xunit}
g_Y^{-2}=\frac{g_5^{-2}}{15}+\frac{g_X^{-2}}{25u^2} ~,
\end{eqnarray}
where $u$ denotes an employed unit of $\rm U(1)_X$ charges. So far
we tacitly supposed $u=1$, but its absolute value should be
determined by the string theory with $g_5=g_X$ at the
compactification scale. Along the standard model direction,
Eq.~(\ref{Xunit}) is still valid above the flipped SU(5) breaking
scale. Thus at the compactification scale, where SU(2) gauge
coupling $g_2$ is identified with $g_5$, the Weinberg angle is given
by
\begin{eqnarray} \label{thetaW}
{\rm sin}^2\theta_W^0=\frac{1}{1+\frac{g_5^2}{g_Y^2}}
=\frac{1}{1+\left(\frac{1}{15}+\frac{1}{25u^2}\right)} ~.
\end{eqnarray}

Including the unit factor `$u$', the $\rm U(1)_X$ charge operator,
$Q^A_X\equiv -2(1,1,1,1,1,0,0,0)(0^8)'\times u$ can be expressed in
terms of a proper orthonomal basis vector $q_5$,
\begin{eqnarray}
Q^A_X=\frac{1}{\sqrt{2}}~c_5q_5 ~,
\end{eqnarray}
where
\begin{eqnarray}
c_5\equiv -2\sqrt{10}~u ~,~~{\rm and}~~ q_5\equiv
\frac{1}{\sqrt{5}}(1,1,1,1,1,0,0,0)(0^8) ~.
\end{eqnarray}
With this expression, $g_5^2/g_X^2$ can be calculated in the string
theory framework~\cite{ChoiKim05},
\begin{eqnarray}
\frac{g_5^2}{g_X^2}=c_5^2=40~u^2 ~.
\end{eqnarray}
Hence, the string theory determines the absolute values of $\rm
U(1)_X$ charges with $u^2=\frac{1}{40}$ from $g_5=g_X$ at the
compactification scale. With this unit, the bare value of the
Weinberg angle is also determined from Eq.~(\ref{thetaW}):
\begin{eqnarray}
{\sin}^2\theta_W^0=\frac{3}{8} ~.
\end{eqnarray}

\subsection{Embedding of $Y$ in flipped SU(5)}

A covariant derivative in flipped SU(5) includes the term
\begin{eqnarray} \label{covD}
\sqrt{\frac{3}{5}}Y_5g_5A_5+X u g_XA_x ~,
\end{eqnarray}
where $Y_5={\rm
diag}.(\frac13,\frac13,\frac13,\frac{-1}2,\frac{-1}2)$ and $X$
[$=X\cdot {\rm diag}.(1,1,1,1,1)$] denotes the $U(1)_X$ charge
operator employed in this paper, $u$ a unit of $U(1)_X$ charge. So
far $u$ is tacitly assumed to be unity, but generically it is not
necessarily 1. $\sqrt{\frac{3}{5}}Y_5$ is one of SU(5) generator
normalized with Tr$T_{SU(5)}T_{SU(5)}=\frac{1}{2}$. $A_5$ and
$A_x$ stand for the SU(5) and $U(1)_X$ gauge fields, respectively.
After symmetry breaking, one linear combination of $A_5$ and $A_x$
becomes the $\rm U(1)_Y$ gauge field $B_\mu$ of the standard
model. We introduce a mixing angle $\phi$,
\begin{eqnarray}
\left(\begin{array}{c} A_5
\\A_x\end{array}\right)=\left(\begin{array}{cc} {\rm cos}\phi &-{\rm sin}\phi
\\{\rm sin}\phi & {\rm cos}\phi
\end{array}\right)\left(\begin{array}{c} B_\mu
\\C_\mu\end{array}\right) ~,
\end{eqnarray}
Eq.~(\ref{covD}) is recast in the following form,
\begin{eqnarray}
\bigg[\frac{1}{5}Y_5+\frac{1}{5}X\sqrt{\frac{5}{3}}\frac{g_X}{g_5}{\rm
tan}\phi ~u\bigg]\sqrt{15}~{\rm cos}\phi ~g_5B_\mu+\bigg[C_\mu~{\rm
terms}\bigg] ~,
\end{eqnarray}
where $C_\mu$ achieves a superheavy mass from the 10 dimensional
Higgs VEVs $\langle \nu^c_H\rangle$, $\langle
\overline{\nu}^c_H\rangle$. From this expression, one can read off
the $U(1)_Y$ charges and coupling,
\begin{eqnarray} \label{BC}
&&Y=\frac{1}{5}Y_5+\frac{1}{5}X\sqrt{\frac{5}{3}}\frac{g_X}{g_5}{\rm
tan}\phi ~u ~,
\\
&&\frac{g_Y}{g_5}=\sqrt{15}~{\rm cos}\phi ~, \label{gY}
\end{eqnarray}
Since $Y(\nu^c)=0$ is the SM direction, Eq.~(\ref{BC}) should be
\begin{eqnarray}
&&Y=\frac{1}{5}Y_5+\frac{1}{5}X ~,
\\
{\rm or}~&&\frac{g_5}{g_X}=\sqrt{\frac{5}{3}}~{\rm tan}\phi ~u
 ~. \label{mixing}
\end{eqnarray}
With Eqs.~(\ref{gY}) and (\ref{mixing}), the relation between
$g_5$, $g_X$, and $g_Y$ is derived:
\begin{eqnarray}
g_Y^{-2}=\frac{g_5^{-2}}{15}+\frac{g_X^{-2}}{25u^2} ~.
\end{eqnarray}

\section{Conclusion}

We constructed a supersymmetric flipped-SU(5) model from a
$\Z_{12-I}$ orbifold compactification. The notable features of the
model are
\begin{itemize}
\item From $E_8$, the only nonabelian group is the needed SU(5).
This is possible only for one shift vector in $\Z_{12-I}$. In this
sense, it is the unique $\Z_{12-I}$ flipped-SU(5) model.
\item Three families are obtained. The third
family is located in twisted sectors while  the first two families
are in the untwisted sector. Separating the third family  from the
first two families enables a mass hierarchy of fermions.
\item There exists a doublet triplet splitting from a kind of the
missing partner mechanism.
\item There results only one pair of Higgs doublets.
\item The $R$-parity is present.
\item Allowed Yukawa couplings can generate a GUT scale VEVs of ${\bf
10}_{1}^H$ and  $\overline{\bf 10}_{\bf -1}^H$ for the GUT breaking
down to the standard model.
\item There exist $Q_{\rm em}=\pm\frac12$ particles. But these form
vectorlike representations and most are removed at the GUT scale.
Thus, $\sin^2\theta_W$ is similar to that of SO(10) GUT.
\end{itemize}

 In this paper, all the relevant
Yukawa couplings are derived from string construction. So far, we
have not encountered any serious phenomenological problem.
Successful Yukawa couplings may be generated by appropriate GUT
scale VEVs of singlet fields which are treated here as free
parameters. Finally, it is expected that a standard model can be
derived from $\Z_{12}$ without going through the intermediate stage
of the flipped SU(5) with features discovered in the present model
\cite{IWKIM}. In a future communication, we will tabulate all
computer-searched $\Z_{12}$ orbifold models \cite{jihun}.

\vskip 1cm \centerline{\bf Acknowledgments} We thank Ji-hun Kim for
numerical checking of the spectra, and thank K.-S. Choi and I.-W.
Kim for useful discussions.  One of us (JEK) also thank the Kabli
Institute for Theoretical Physics for the hospitality during this
work was finished.  This research was  supported in part by the
National Science Foundation under Grant No. PHY99-07949. JEK is also
supported in part by the KRF grants, No. R14-2003-012-01001-0, No.
R02-2004-000-10149-0, and No. KRF-2005-084-C00001.



\vskip 1cm \centerline{\bf Appendix: Flipped SU(5) with
SU(4)$^\prime$}\vskip 0.5cm

One can eliminate some exotic particles carrying $Q_X=\pm
\frac12$, $\pm \frac52$ observed in the model discussed in the
main text (Model I) by employing more complicated shift vector and
Wilson line,
\begin{eqnarray}
V&=&\textstyle\left(\frac{1}{4},\frac{1}{4},\frac{1}{4},\frac{1}{4},\frac{1}{4};
\frac{5}{12},\frac{6}{12},0\right)\left(\frac{2}{12},\frac{2}{12};1,0;0^4
\right)' ~,
\\
a_3&=&\textstyle\left(0,0,0,0,0;0,\frac{-1}{3},\frac{1}{3}
\right)\left(0,0;0,0;\frac{2}{3},\frac{2}{3},\frac{-1}{3},\frac{-1}{3}
\right)'~,
\end{eqnarray}
which satisfy the modular invariance conditions.  Model II,
constructed with these shift vector and Wilson line, eliminates in
particular all ${\bf 5}_{1/2}$ and ${\bf \overline{5}}_{-1/2}$ from
the massless spectrum, leaving intact the MSSM fields obtained in
Model I. Here, the gauge group is further broken down to
\begin{eqnarray}
\left[\left\{SU(5)\times U(1)_X\right\}\times
U(1)^3\right]\times\left[SU(4)\times SU(2)_1\times SU(2)_2\times
SU(2)_3\times U(1)^2\right]'~.
\end{eqnarray}
Model II gives the same spectrum as Model I for the $T_6$, $T_2^0$,
and $T_4^0$ sectors and visible sector of $U$ in Model I. As in
Model I, there is no massless states satisfying $(P+3V)\cdot a_3=0$
mod. Z in the $T_3$ sector. The spectrum of Model II is summarized
as follows:

$\bullet$ Fields of Flipped SU(5) :  $3\times {\bf 16_{\rm
flip}}+1\times\{{\bf 5}_{-2},{\bf \overline{5}}_{2},\}+1\times
\{{\bf 10}_1,{\bf \overline{10}}_{-1}\}$,

where ${\bf 16_{\rm flip}}\equiv \{{\bf 10}_1,{\bf
\overline{3}}_{-3},{\bf 1}_5\}$.

$\bullet$ (regularly charged) vector-like fields : $2\times \{{\bf
16_{\rm flip}},{\bf \overline{16}_{\rm flip}}\}+2\times {\bf
10}_{\rm flip}$,

where ${\bf 10_{\rm flip}}\equiv \{{\bf 5}_{-2},{\bf
\overline{5}}_{2}\}$.

$\bullet$ Exotic particles : $16\times \{{\bf 1}_{5/2},{\bf
1}_{-5/2} \}$.

$\bullet$ A lot of neutral singlets under $SU(5)\times U(1)_X$.

\noindent The full massless spectrum of Model II is presented in
the following tables.

\begin{table}
\begin{center}
\begin{tabular}{|c|c|c|c|}
\hline $P\cdot V$& $\tilde s,\ U_i$ & Visible states & $SU(5)\times
U(1)_X$
\\
\hline  & &$(\underline{+----};+++)$ & ${\bf {5}}_{3}^{L}$

\\
$\frac{1}{12}$ & $(+~-~+),\ U_3$ &$(\underline{+++--};+--)$ &
$\overline{\bf 10}_{-1}^{L}$
\\
$$ &  &$(\underline{+++++};+++)$ & ${\bf 1}_{-5}^{L}$
\\
\hline $\frac{4}{12}$ & $(+~+~-),\ U_2$
&$(\underline{-1,0,0,0,0};-1,0,0)$ & $\overline{\bf 5}_{2}^{L}$
\\
\hline &  & $(\underline{++++-};-++)$ & $\overline{\bf
{5}}_{-3}^{R}$
\\
$\frac{5}{12}$&  $(+~+~+),\ U_1$ & $(\underline{++---};---)$ & ${\bf
10}_{1}^{R}$
\\
$$ & &$(-----;-++)$ & ${\bf 1}_{5}^{R}$
\\
\hline
\hline $P\cdot V$ & $\tilde s,\ U_i$ & Hidden states &
$[SU(4)\times SU(2)^3]'$
\\
\hline
 &  &$(--;++;\pm\pm\mp\mp)'$, $(--;--;\pm\pm\mp\mp)'$ & $$
\\
$\frac{4}{12}$  & $(+~+~-),\ U_2$ &$(--;++;\pm\mp\pm\mp)'$,
$(--;--;\pm\mp\pm\mp)'$ & $({\bf 6},{\bf 2},{\bf 1},{\bf 1})'_L$
\\
 &  &$(--;++;\pm\mp\mp\pm)'$, $(--;--;\pm\mp\mp\pm)'$ & $$
\\
\hline $\frac{4}{12}$& $(+~+~-),\ U_2$ & $(1,1;0,0;0,0,0,0)'$ &
singlet
\\
\hline
\end{tabular}
\end{center}
\caption{Chiral fields from the $U$-sector. $+$ ($-$) denotes
$\frac{1}{2}$ ($\frac{-1}{2}$). }\label{Atb:untwistedhd}
\end{table}

\begin{table}
\begin{center}
\begin{tabular}{c|c|c|c|c}
\hline  $P+6V$ & $(N^L)_j$ & $\Theta_0$ & ${\cal P}_6$& $\chi$
\\
\hline
 $\left({\bf 5}_{3};\frac{1}{2},0,0\right)(0^8)'$ & $0$ & $\frac{1}{2}$ & $2$
&$L$
\\
 $\left(\overline{{\bf 10}}_{-1};\frac{1}{2},0,0\right)(0^8)'$ & $0$ &
$0$ & $4$ &$L$
\\
$\left({\bf 1}_{-5};\frac{1}{2},0,0\right)(0^8)'$ & $0$ &
$\frac{1}{2}$ & $2$ &$L$
\\
\hline
 $\left(\overline{{\bf 5}}_{-3};\frac{-1}{2},0,0\right)(0^8)'$ & $0$ &
 $\frac{-1}{6}$ & $2$ &$L$
\\
$\left(
 {\bf 10}_{1};\frac{-1}{2},0,0\right)(0^8)'$ & $0$ & $\frac{1}{3}$ & $3$
&$L$
\\
 $\left({\bf{1}}_{5};\frac{-1}{2},0,0\right)(0^8)'$ & $0$ &
$\frac{-1}{6}$ & $2$ &$L$
\\ \hline
$$ & $1_3$ & $0$ & $4$ &$L$
\\
$\left({\bf 1}_0;0,\frac{1}{2},\frac{1}{2}\right)(0^8)'$ &
$1_{\bar{3}}$ & $\frac{-1}{6}$ & $2$ &$L$
\\
$$ & $1_{\bar{1}}$ & $\frac{1}{2}$ & $2$ &$L$
\\
$$ & $1_{1}$ & $\frac{1}{3}$ & $3$ &$L$
\\
\hline
$$ & $1_3$ & $\frac{1}{2}$ & $2$ &$L$
\\
$\left({\bf 1}_0;0,\frac{-1}{2},\frac{-1}{2}\right)(0^8)'$ &
$1_{\bar{3}}$ & $\frac{1}{3}$ & $3$ &$L$
\\
$$ & $1_{\bar{1}}$ & $0$ & $4$ &$L$
\\
$$ & $1_{1}$ & $\frac{-1}{6}$ & $2$ &$L$
\\
\hline
\end{tabular}
\end{center}
\caption{Massless states satisfying $(P+6V)\cdot a_3=0 $ mod. Z in
$T_6$. The definitions of ${\bf 5}_3$, ${\bf \overline{5}}_{-3}$,
${\bf 10}_1$, ${\bf \overline{10}}_{-1}$, and ${\bf 1}_{\pm 5,0}$
are found in the main text.} \label{Atb:T6states}
\end{table}

\begin{table}
\begin{center}
\begin{tabular}{c|c|c|c}
\hline $P+2V$ & $(N^L)_j$ & ${\cal P}_2(f_0)$ &$\chi$
\\
\hline
$\left({\bf{5}}_{3};\frac{-1}{6},0^2\right)(\frac{1}{3},\frac{1}{3};0^6)'$
& $0$ & $1$ &$L$
\\
$\left({\bf
1}_{-5};\frac{-1}{6},0^2\right)(\frac{1}{3},\frac{1}{3};0^6)'$ & $0$
& $1$ &$L$
\\
\hline $\left({\bf
1}_0;\frac{1}{3},\frac{1}{2},\frac{-1}{2}\right)(\frac{1}{3},\frac{1}{3};0^6)'$
& $2_{\bar{1}},~2_3$ & $1+1$ &$L$
 \\
$\left({\bf
1}_0;\frac{1}{3},\frac{-1}{2},\frac{1}{2}\right)(\frac{1}{3},\frac{1}{3};0^6)'$
& $1_{\bar{2}},\{1_{\bar{1}}+1_3\}$ & $1+1$ &$L$
\\
$\left({\bf
1}_0;\frac{-2}{3},\frac{1}{2},\frac{1}{2}\right)(\frac{1}{3},\frac{1}{3};0^6)'$
& $1_{\bar{1}}$ & $1$ &$L$
\\
$\left({\bf
1}_0;\frac{-2}{3},\frac{-1}{2},\frac{-1}{2}\right)(\frac{1}{3},\frac{1}{3};0^6)'$
& $1_3$ & $1$ &$L$
\\
$\left({\bf
1}_0;\frac{1}{3},\frac{-1}{2},\frac{1}{2}\right)(\frac{-2}{3},\frac{-2}{3};0^6)'$
& $0$ & $1$ &$L$
\\
\hline \hline $P+2V_+$ & $(N^L)_j$ & ${\cal P}_2(f_+)$& $\chi$
\\
\hline $\left({\bf
1}_0;\frac{1}{3},\frac{-1}{6},\frac{1}{6}\right)({\bf
\overline{4}},{\bf 1},{\bf 1},{\bf 2})_{2+}'$ & $0$ & $1$ & $L$
\\
$\left({\bf 1}_0;\frac{1}{3},\frac{-1}{6},\frac{1}{6}\right)({\bf
1},{\bf 1},{\bf 2},{\bf 2})_{2+}'$ & $0$ & $1$ & $L$
\\
$\left({\bf 1}_0;\frac{1}{3},\frac{-1}{6},\frac{1}{6}\right)({\bf
1},{\bf 2},{\bf 1},{\bf 1})_{2+}'$ & $2_{\bar{1}},~2_3$ & $1+1$ &
$L$
\\
$\left({\bf 1}_0;\frac{1}{3},\frac{-1}{6},\frac{1}{6}\right)({\bf
1},{\bf 1},{\bf 1},{\bf 1})_{2+}'$ & $2_{\bar{1}},~ 2_3$ & $1+1$ &
$L$
\\
\hline \hline $P+2V_-$ & $(N^L)_j$ & ${\cal P}_2(f_-)$& $\chi$
\\
\hline $\left({\bf
1}_0;\frac{1}{3},\frac{1}{6},\frac{-1}{6}\right)({\bf 6},{\bf
1},{\bf 1},{\bf 1})_{2-}'$ & $0$ & $1$ & $L$
\\
$\left({\bf 1}_0;\frac{1}{3},\frac{1}{6},\frac{-1}{6}\right)({\bf
\overline{4}},{\bf 1},{\bf 2},{\bf 1})_{2-}'$ & $0$ & $1$& $L$
\\
$\left({\bf 1}_0;\frac{1}{3},\frac{1}{6},\frac{-1}{6}\right)({\bf
1},{\bf 1},{\bf 1},{\bf 1})_{2-}'$ & $0$ & $1$& $L$
\\
$\left({\bf 1}_0;\frac{1}{3},\frac{1}{6},\frac{-1}{6}\right)({\bf
1},{\bf 2},{\bf 1},{\bf 1})_{2-}'$ &
$1_{\bar{2}},~\{1_{\bar{1}}+1_3\}$ & $1+1$& $L$
\\
$\left({\bf 1}_0;\frac{1}{3},\frac{1}{6},\frac{-1}{6}\right)({\bf
1},{\bf 1},{\bf 1},{\bf 1})_{2-}''$ &
$1_{\bar{2}},~\{1_{\bar{1}}+1_3\}$ & $1+1$& $L$
\\
\hline
\end{tabular}
\end{center}
\caption{Chiral matter fields satisfying $\Theta_{0,+,-}=0$ in the
$T_2^{0,+,-}$ sectors.}\label{Atb:T2-}
\end{table}

\begin{table}
\begin{center}
\begin{tabular}{c|c|c|c|c}
\hline $P+4V$ & $(N^L)_j$ &
 $\Theta_0$ & ${\cal P}_4(f_0)$ &$\chi$
\\
\hline $\left({\bf{5}}_{-2};\frac{-1}{3},0^2\right)
(\frac{-1}{3},\frac{-1}{3};0^6)'$ & $0$ & $0$ & $3$ &$L$
\\
$\left(\overline{\bf 5}_{2}
;\frac{-1}{3},0^2\right)(\frac{-1}{3},\frac{-1}{3};0^6)'$ & $0$ &
$\frac{1}{2}$ & $2$&$L$
\\
\hline
$$ & $1_{\bar{1}}$ & $\frac{-1}{4}$ & $2$&$L$
\\
$\left({\bf
1}_0;\frac{2}{3},0^2\right)(\frac{-1}{3},\frac{-1}{3};0^6)'$ & $1_2$
& $\frac{1}{2}$ & $2$&$L$
\\
$$ & $1_3$ & $\frac{1}{4}$ & $2$&$L$
\\
\hline $\left({\bf
1}_0;\frac{2}{3},0^2\right)(\frac{2}{3},\frac{2}{3};0^6)'$ & $0$ &
$\frac{1}{2}$ & $2$&$L$
\\
$\left({\bf 1}_0;\frac{-1}{3},\pm
1,0\right)(\frac{-1}{3},\frac{-1}{3};0^6)'$ & $0$ & $\frac{1}{4}$ &
$2+2$&$L$
\\
$\left({\bf 1}_0;\frac{-1}{3},0,\pm
1\right)(\frac{-1}{3},\frac{-1}{3};0^6)'$ & $0$ & $\frac{-1}{4}$ &
$2+2$&$L$
\\
\hline\hline $P+4V_+$ & $(N^L)_j$ & $\Theta_+$ & ${\cal P}_4(f_+)$
&$\chi$
\\
\hline $\left({\bf
1}_0;\frac{2}{3},\frac{-1}{3},\frac{1}{3}\right)({\bf 1},{\bf
2},{\bf 1},{\bf 1})_{4+}'$ & $0$ & $0$ & $3$&$L$
\\
$\left({\bf 1}_0;\frac{2}{3},\frac{-1}{3},\frac{1}{3}\right)({\bf
1},{\bf 1},{\bf 1},{\bf 1})_{4+}'$ & $0$ & $0$ & $3$&$L$
\\
\hline $\left({\bf
1}_0;\frac{-1}{3},\frac{-1}{3},\frac{-2}{3}\right)({\bf 1},{\bf
2},{\bf 1},{\bf 1})_{4+}'$ & $0$ & $\frac{1}{4}$ & $2$&$L$
\\
$\left({\bf 1}_0;\frac{-1}{3},\frac{-1}{3},\frac{-2}{3}\right)({\bf
1},{\bf 1},{\bf 1},{\bf 1})_{4+}'$ & $0$ & $\frac{1}{4}$ & $2$&$L$
\\
\hline $\left({\bf
1}_0;\frac{-1}{3},\frac{2}{3},\frac{1}{3}\right)({\bf 1},{\bf
2},{\bf 1},{\bf 1})_{4+}'$ & $0$ & $\frac{-1}{4}$ & $2$&$L$
\\
$\left({\bf 1}_0;\frac{-1}{3},\frac{2}{3},\frac{1}{3}\right)({\bf
1},{\bf 1},{\bf 1},{\bf 1})_{4+}'$ & $0$ & $\frac{-1}{4}$ & $2$&$L$
\\
\hline \hline $P+4V_-$ & $(N^L)_j$ & $\Theta_-$ & ${\cal
P}_4(f_-)$&$\chi$
\\
\hline $\left({\bf
1}_0;\frac{2}{3},\frac{1}{3},\frac{-1}{3}\right)({\bf 1},{\bf
2},{\bf 1},{\bf 1})_{4-}'$ & $0$ & $0$ & $3$&$L$
\\
$\left({\bf 1}_0;\frac{2}{3},\frac{1}{3},\frac{-1}{3}\right)({\bf
1},{\bf 1},{\bf 1},{\bf 1})_{4-}'$ & $0$ & $0$ & $3$&$L$
\\
\hline $\left({\bf
1}_0;\frac{-1}{3},\frac{1}{3},\frac{2}{3}\right)({\bf 1},{\bf
2},{\bf 1},{\bf 1})_{4-}'$ & $0$ & $\frac{1}{4}$ & $2$&$L$
\\
$\left({\bf 1}_0;\frac{-1}{3},\frac{1}{3},\frac{2}{3}\right)({\bf
1},{\bf 1},{\bf 1},{\bf 1})_{4-}'$ & $0$ & $\frac{1}{4}$ & $2$&$L$
\\
\hline $\left({\bf
1}_0;\frac{-1}{3},\frac{-2}{3},\frac{-1}{3}\right)({\bf 1},{\bf
2},{\bf 1},{\bf 1})_{4-}'$ & $0$ & $\frac{-1}{4}$ & $2$&$L$
\\
$\left({\bf 1}_0;\frac{-1}{3},\frac{-2}{3},\frac{-1}{3}\right)({\bf
1},{\bf 1},{\bf 1},{\bf 1})_{4-}'$ & $0$ & $\frac{-1}{4}$ & $2$&$L$
\\
\hline
\end{tabular}
\end{center}
\caption{Chiral matter fields in the $T_4^0, T_4^+$ and $T_4^-$
sectors. In the $T_4^+$ and $T_4^-$ sectors, $({\bf 1},{\bf 2},{\bf
1},{\bf 1})_{4+}'\equiv \left(\frac{1}{6},\frac{1}{6};\pm\pm;
\frac{1}{6},\frac{1}{6},\frac{1}{6},\frac{1}{6}\right)'$,  $({\bf
1},{\bf 1},{\bf 1},{\bf 1})_{4+}'\equiv
\left(\frac{-1}{3},\frac{-1}{3};0,0;
\frac{-1}{3},\frac{-1}{3},\frac{-1}{3},\frac{-1}{3}\right)'$, $({\bf
1},{\bf 2},{\bf 1},{\bf 1})_{4-}'=
\left(\frac{1}{6},\frac{1}{6};\pm\pm;
\frac{-1}{6},\frac{-1}{6},\frac{-1}{6},\frac{-1}{6}\right)'$ and
$({\bf 1},{\bf 1},{\bf 1},{\bf 1})_{4-}'=
\left(\frac{-1}{3},\frac{-1}{3};0,0;
\frac{1}{3},\frac{1}{3},\frac{1}{3},\frac{1}{3}\right)'$.}\label{Atb:T4}
\end{table}

\begin{table}
\begin{center}
\begin{tabular}{c|c|c|c}
\hline $P+V$ & $(N^L)_j$ & ${\cal P}_1(f_0)$ &$\chi$
\\
\hline $\left({\bf 1}_{-5/2};\frac{5}{12},\frac{6}{12},0\right)({\bf
1},{\bf 1},{\bf 1},{\bf 2})_{1}'$ & $0$ & $1$ &$L$
\\
$\left({\bf 1}_{5/2};\frac{-1}{12},0,\frac{-6}{12}\right)({\bf
1},{\bf 1},{\bf 1},{\bf 2})_{1}'$ & $1_3$ & $1$ &$L$
\\
\hline \hline $P+V_+$ & $(N^L)_j$ & ${\cal P}_1(f_+)$&$\chi$
\\
\hline $\left({\bf
1}_{-5/2};\frac{5}{12},\frac{2}{12},\frac{4}{12}\right)({\bf 4},{\bf
1},{\bf 1},{\bf 1})_{1+}'$ & $0$ & $1$ &$L$
\\
$\left({\bf
1}_{-5/2};\frac{5}{12},\frac{2}{12},\frac{4}{12}\right)({\bf 1},{\bf
1},{\bf 2},{\bf 1})_{1+}'$ & $0$ & $1$ &$L$
\\
$\left({\bf
1}_{5/2};\frac{-1}{12},\frac{-4}{12},\frac{-2}{12}\right)({\bf
4},{\bf 1},{\bf 1},{\bf 1})_{1+}'$ & $1_3$ & $1$ &$L$
\\
$\left({\bf
1}_{5/2};\frac{-1}{12},\frac{-4}{12},\frac{-2}{12}\right)({\bf
1},{\bf 1},{\bf 2},{\bf 1})_{1+}'$ & $1_3$ & $1$ &$L$
\\
\hline
\hline  $P+5V$ & $(N^L)_j$ & ${\cal P}_5(f_0)$ &$\chi$
\\
\hline   $\left({\bf
1}_{5/2};\frac{1}{12},\frac{-6}{12},0\right)({\bf 1},{\bf 1},{\bf
1},{\bf 2})_{5}'$ & $1_{1}$ & $1$&$R$
\\
 $\left({\bf 1}_{-5/2};\frac{-5}{12},0,\frac{6}{12}\right)({\bf
1},{\bf 1},{\bf 1},{\bf 2})_{5}'$ & $0$ & $1$&$R$
\\
\hline \hline $P+5V_+$ & $(N^L)_j$ & ${\cal P}_5(f_+)$ &$\chi$
\\
\hline   $\left({\bf
1}_{5/2};\frac{1}{12},\frac{-2}{12},\frac{-4}{12}\right)({\bf
\overline{4}},{\bf 1},{\bf 1},{\bf 1})_{5+}'$ & $1_{1}$ & $1$ &$R$
\\
 $\left({\bf
1}_{5/2};\frac{1}{12},\frac{-2}{12},\frac{-4}{12}\right)({\bf
1},{\bf 1},{\bf 2},{\bf 1})_{5+}'$ & $1_{1}$ & $1$ &$R$
\\
 $\left({\bf
1}_{-5/2};\frac{-5}{12},\frac{4}{12},\frac{2}{12}\right)({\bf
\overline{4}},{\bf 1},{\bf 1},{\bf 1})_{5+}'$ & $0$ & $1$ &$R$
\\
 $\left({\bf
1}_{-5/2};\frac{-5}{12},\frac{4}{12},\frac{2}{12}\right)({\bf
1},{\bf 1},{\bf 2},{\bf 1})_{5+}'$ & $0$ & $1$ &$R$
\\
\hline
\end{tabular}
\end{center}
\caption{Chiral matter fields satisfying $\Theta_{0,+}=0$ in the
$T_1^0$, $T_1^+$, $T_5^0$, and $T_5^+$ sectors.  There are no
massless states in $T_1^-$ and $T_5^-$.  Here, ${\bf 1}_{\mp
5/2}\equiv
(\pm\frac{1}{4},\pm\frac{1}{4},\pm\frac{1}{4},\pm\frac{1}{4},\pm\frac{1}{4}
)$. In $T_1^0$ and $T_1^+$, $({\bf 4},{\bf 1},{\bf 1},{\bf
1})_{1+}' \equiv ( \frac{1}{6},\frac{1}{6};0,0;
\underline{\frac{2}{3},\frac{-1}{3},\frac{-1}{3},\frac{-1}{3}})'
$, $({\bf 1},{\bf 1},{\bf 2},{\bf 1})_{1+}' \equiv (
\frac{-1}{3},\frac{-1}{3};
\pm\mp;\frac{1}{6},\frac{1}{6},\frac{1}{6},\frac{1}{6})' $.
In $T_5^{0}$ and $T_5^+$,
$({\bf 1},{\bf 1},{\bf 1},{\bf 2})_{5}'=
\left(\underline{\frac{5}{6},\frac{-1}{6}};0,0,;0,0,0,0 \right)'
$, $({\bf \overline{4}},{\bf 1},{\bf 1},{\bf 1})_{5+}' = \left(
\frac{-1}{6},\frac{-1}{6};0,0;
\underline{\frac{-2}{3},\frac{1}{3},\frac{1}{3},\frac{1}{3}}\right)'$,
and $({\bf 1},{\bf 1},{\bf 2},{\bf 1})_{5+}' = \left(
\frac{1}{3},\frac{1}{3};
\pm\mp;\frac{-1}{6},\frac{-1}{6},\frac{-1}{6},\frac{-1}{6}\right)'$.
}\label{Atb:T10}
\end{table}

In Table \ref{Atb:T2-}, the abbreviated symbols denote
\begin{eqnarray}
({\bf \overline{4}},{\bf 1},{\bf 1},{\bf 2})_{2+}'&\equiv&\textstyle
\left(\underline{\frac{-2}{3},\frac{1}{3}};0,0;
\underline{\frac{-2}{3},\frac{1}{3},\frac{1}{3},\frac{1}{3}}\right)'
~,
\nonumber \\
({\bf 1},{\bf 1},{\bf 2},{\bf 2})_{2+}'&\equiv&\textstyle
\left(\underline{\frac{5}{6},\frac{-1}{6}};\underline{\frac{1}{2},\frac{-1}{2}};
\frac{-1}{6},\frac{-1}{6},\frac{-1}{6},\frac{-1}{6}\right)' ~,
\nonumber \\
({\bf 1},{\bf 2},{\bf 1},{\bf 1})_{2+}'&\equiv&\textstyle
\left(\frac{-1}{6},\frac{-1}{6};\pm\frac{1}{2},\pm\frac{1}{2}
;\frac{-1}{6},\frac{-1}{6},\frac{-1}{6},\frac{-1}{6}\right)' ~,
\nonumber \\
({\bf 1},{\bf 1},{\bf 1},{\bf 1})_{2+}'&\equiv&\textstyle
\left(\frac{1}{3},\frac{1}{3};0,0;\frac{1}{3},\frac{1}{3},
\frac{1}{3},\frac{1}{3}\right)' ~,
\nonumber \\
({\bf 6},{\bf 1},{\bf 1},{\bf 1})_{2-}'&\equiv&
\left\{\begin{array}{c}
\left(\frac{1}{3},\frac{1}{3};0,0,;\frac{2}{3},\frac{2}{3},\frac{-1}{3},\frac{-1}{3}
\right)'\nonumber
\\
\left(\frac{1}{3},\frac{1}{3};0,0,;\frac{-1}{3},\frac{-1}{3},\frac{2}{3},\frac{2}{3}
\right)'\nonumber
\\
\left(\frac{1}{3},\frac{1}{3};0,0,;\underline{\frac{2}{3},\frac{-1}{3}},
\underline{\frac{2}{3},\frac{-1}{3}} \right)'
\end{array}\right.\nonumber
\\
({\bf \overline{4}},{\bf 1},{\bf 2},{\bf 1})_{2-}'&\equiv&\textstyle
\left(\frac{-1}{6},\frac{-1}{6};\pm\mp ;
\underline{\frac{-5}{6},\frac{1}{6},\frac{1}{6},\frac{1}{6}}\right)'\nonumber
\\
({\bf 1},{\bf 1},{\bf 1},{\bf 1})_{2-}'&\equiv&\textstyle
\left(\frac{-2}{3},\frac{-2}{3};0,0
;\frac{-1}{3},\frac{-1}{3},\frac{-1}{3},\frac{-1}{3}\right)'\nonumber
\\
({\bf 1},{\bf 2},{\bf 1},{\bf 1})_{2-}'&\equiv&\textstyle
\left(\frac{-1}{6},\frac{-1}{6};\pm\pm
;\frac{1}{6},\frac{1}{6},\frac{1}{6},\frac{1}{6}\right)'\nonumber
\\
({\bf 1},{\bf 1},{\bf 1},{\bf 1})_{2-}''&\equiv&\textstyle
\left(\frac{1}{3},\frac{1}{3};0,0;\frac{-1}{3},
\frac{-1}{3},\frac{-1}{3},\frac{-1}{3}\right)'\nonumber
\end{eqnarray}

\newpage

\end{document}